\documentclass[aps,prl,twocolumn,showpacs,preprintnumbers,amsmath,amssymb,superscriptaddress,longbibliography,nofootinbib]{revtex4-2}
\usepackage{hyperref}
\hypersetup{colorlinks=true, citecolor=blue, linkcolor=blue,urlcolor=cyan}
\usepackage{color}
\usepackage{graphicx}
\usepackage{dcolumn}
\usepackage{braket}
\usepackage{amsthm}
\usepackage[caption=false]{subfig}
\usepackage{booktabs}

\newcommand{\kett}[1]{ \ket{#1}\!\rangle }
\newcommand{\suk}[1]{\widehat{\mathfrak{su}}(2)_{#1}}

\makeatletter
\let\cat@comma@active\@empty
\makeatother

\begin{document}
\preprint{KYUSHU-HET-392}
\preprint{RIKEN-iTHEMS-Report-26}
\title{
Identifying Renormalization Group Interfaces of
\\
Massless WZW Flows \(\widehat{\mathfrak{su}}(2)_k \to \widehat{\mathfrak{su}}(2)_1\) 
}

\author{Yuma Furuta }\email[]{furuta.yuma@phys.kyushu-u.ac.jp}
\affiliation{\it Institute for Advanced Study, Kyushu University, Fukuoka 819-0395, Japan.}
\author{Yuya Kusuki}\email[]{kusuki.yuya@phys.kyushu-u.ac.jp}
\affiliation{\it Institute for Advanced Study, Kyushu University, Fukuoka 819-0395, Japan.}
\affiliation{\it Department of Physics, Kyushu University, Fukuoka 819-0395, Japan.}
\affiliation{\it RIKEN Interdisciplinary Theoretical and Mathematical Sciences (iTHEMS), Wako, Saitama 351-0198, Japan.}
\affiliation{\it JST, PRESTO, 4-1-8 Honcho, Kawaguchi, Saitama 332-0012, Japan.}
\author{Toshiki Onagi}\email[]{toshiki.onagi@yukawa.kyoto-u.ac.jp}
\affiliation{\it Yukawa Institute for Theoretical Physics, Kyoto University, Kitashirakawa Oiwakecho, Sakyo-ku, Kyoto 606-8502, Japan.}
\author{Yin Tang }\email[]{tang.yin.129@m.kyushu-u.ac.jp}
\affiliation{\it Institute for Advanced Study, Kyushu University, Fukuoka 819-0395, Japan.}

\begin{abstract}
A renormalization-group interface encodes how a flow relates operators at short and long distances. We identify this interface up to topological dressing for the conjectured massless \(\widehat{\mathfrak{su}}(2)_k \to \widehat{\mathfrak{su}}(2)_1\) Wess–Zumino–Witten flows at every odd level \(k\geq 3\). Global \(SU(2)\) symmetry requires the interface to preserve a diagonal current algebra at level \(k+1\), built from the currents on its two sides. The \(\mathbb Z_2\) center symmetry then selects which representation labels the interface. This identification predicts distinct patterns of low-lying interface excitations for flows to the same infrared fixed point. Spin-chain calculations reproduce these patterns for \(k=3\) and \(5\), providing independent evidence for both conjectured flows.
\end{abstract}
\maketitle

\textit{Introduction.}---Renormalization-group (RG) interfaces encode how two critical theories are connected along a flow~\cite{BrunnerRoggenkamp2008,Gaiotto:2012np}. They arise when a perturbation acts on only one half of space. Folding turns the interface into a boundary condition for the product of the ultraviolet \mbox{(UV)} and infrared \mbox{(IR)} theories~\cite{Oshikawa1996,Oshikawa1997,Bachas2002,PetkovaZuber2001,FFRS2004,FFRS2007,BachasBrunner2008,Wong1993,Cardy1989,Ishibashi1989,RecknagelSchomerus2013}. Proposed RG interfaces for minimal models and related coset theories~\cite{Gaiotto:2012np,Poghosyan2014,Stanishkov2016a,Poghosyan2022W3} have been tested by conformal perturbation theory~\cite{Poghossian2014,Poghosyan2013mixing,AhnStanishkov2014,Stanishkov2016b,Poghosyan2023A2} and, for the flow from tricritical Ising to Ising, by lattice calculations of correlation functions~\cite{Cogburn2024} and excitation spectra~\cite{AntunesRong:2026w3isingtim}. See also work on RG interfaces~\cite{AntunesChenRong2026,Zhang2026,Chen2026a} and symmetry constraints on bulk flows~\cite{TanakaNakayama2024,AmbrosinoNegro2025,BenedettiFendleyMagan2026}.

Here we study the RG interfaces of the conjectured massless flows between Wess--Zumino--Witten (WZW) models,
\begin{equation}
  \suk{k}\longrightarrow\suk{1},\qquad k\ge3\ \text{odd}.
  \label{eq:flow}
\end{equation}
The physical motivation comes from Haldane's conjecture that antiferromagnetic Heisenberg chains are gapped for integer spin and gapless for half-integer spin~\cite{Haldane1983,AffleckLieb1986}. In the field-theory description of the half-integer case~\cite{Affleck1985,Affleck1986NPB,Affleck1987}, a spin-$S$ chain can be tuned to a $\suk{2S}$ critical point~\cite{Michaud:2013nq,chepiga2020floating,Chepiga:2022ciu}. Perturbing this point toward the gapless phase is expected to lead to the same $\suk{1}$ IR theory for all half-integer $S>1/2$~\cite{Affleck1987,Konik2015,James2018,Lecheminant2015}.

In this letter, we identify the RG interface up to topological dressing for every flow in Eq.~\eqref{eq:flow}. Locality, unitarity, and global $\mathrm{SU}(2)$ symmetry force the folded interface to preserve the diagonal $\suk{k+1}$ current algebra. The coset is a Virasoro minimal model, so conformal invariance also fixes the gluing of its full chiral algebra. We show that the known coset boundary states~\cite{QuellaSchomerus2002,Quella:2006de} exhaust the elementary possibilities. The $\mathbb Z_2$ center symmetry then selects a single interface up to topological dressing.

\begin{figure}[t]
\centering
\includegraphics[width=0.48\textwidth]{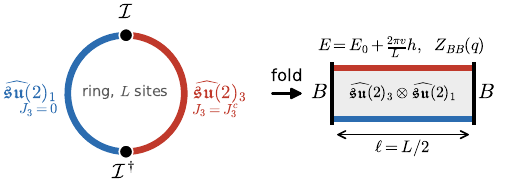}
\caption{A periodic chain whose left half is in the $\suk{1}$ phase and whose right half is tuned to the $\suk{3}$ point contains two RG interfaces $\mathcal I$, $\mathcal I^\dagger$. A reflection exchanges the two junctions while preserving each half, so both fold to the same boundary state $B$; for equal propagation times the ring is described by a strip of width $\ell=L/2$ with conformal spectrum $Z_{BB}$.}
\label{fig:geometry}
\end{figure}

Although the flows share an IR fixed point, the ratio of the lowest weights in different $S^z$ sectors is generically distinguished for different $k$. At fixed $k$, the energy-transmission coefficient and the lowest conformal weight in each magnetization sector are independent of the topological dressing. To test these predictions, we perturb one segment of a periodic spin chain, producing two interfaces (Fig.~\ref{fig:geometry}). When the two segments have equal propagation times, ratios of these weights give the corresponding finite-size gap ratios. The spin-$3/2$ and spin-$5/2$ chains reproduce the gap patterns predicted for $k=3$ and $k=5$, respectively. The first excitation in the zero-magnetization sector provides a separate test of the dressing and disfavors the non-invertibly dressed candidates in both chains. These lattice spectra support the interface identification and provide evidence for the conjectured WZW flows.

\textit{Identifying the RG interface from symmetry.}---For odd $k\ge3$, consider a conformal interface between the $\suk{k}$ and $\suk{1}$ WZW models. Folding gives a boundary of the product theory $\suk{k}\otimes\suk{1}$. We denote its boundary state by $\ket B$ and assume that the boundary CFT is local and unitary and that the boundary condition is elementary, meaning that the identity is its only boundary field of dimension zero. The UV adjoint perturbation $\Phi^{(1)}=\mathrm{tr}_{\rm adj}\,g$, of dimension $4/(k+2)$, preserves two symmetries. Global $\mathrm{SU}(2)$ rotations act on the WZW field $g$ by conjugation, with faithful action $\mathrm{SO}(3)$. The distinct $\mathbb Z_2$ center acts by $g\mapsto-g$, multiplying spin-$j$ primaries by $(-1)^{2j}$, and represents one-site translation in the continuum limit of the homogeneous chain~\cite{Affleck1987,Michaud:2013nq,chepiga2020floating}. We ask which elementary boundaries preserve both symmetries when each acts simultaneously on the two factors.

Consider first the spin rotations. Their holomorphic current is the sum $J^a_{\rm diag}=J^a_{(k)}+J^a_{(1)}$ of the currents of the two folded theories, with an analogous antiholomorphic current $\bar J^a_{\rm diag}$. On the upper half plane the normal current is $j^a_y=i(J^a_{\rm diag}-\bar J^a_{\rm diag})$. Local charge conservation relates the normal current to the change in any charge $j^a_\partial$ localized at the boundary through the Ward identity $j^a_y(\tau,0)=-\partial_\tau j^a_\partial(\tau)$~\cite{ThorngrenWang2021,JensenShaverinYarom2018}. Since the bulk current has scaling dimension one, the Ward identity requires the boundary operator $j^a_\partial$ to have scaling dimension $h=0$. Positivity of the spectrum then gives $L_1\ket{j^a_\partial}=0$, and unitarity makes its derivative null since $\|L_{-1}\ket{j^a_\partial}\|^2=2h\,\|\ket{j^a_\partial}\|^2=0$. Thus the normal current vanishes, and the holomorphic and antiholomorphic currents agree at the boundary. Equivalently, on the disk the Laurent modes obey
\begin{equation}
  \big(J^a_{{\rm diag},n}+\bar J^a_{{\rm diag},-n}\big)\ket{B}=0.
  \label{eq:gluing}
\end{equation}
The levels of the two commuting current algebras add, so this gluing preserves the diagonal $\suk{k+1}$ algebra.

At a conformal boundary the holomorphic and antiholomorphic components of the total stress tensor agree. For the folded theory this means $T_{(k)}+T_{(1)}=\bar T_{(k)}+\bar T_{(1)}$. The Sugawara stress tensor $T_{\rm Sug}$ is quadratic in the diagonal currents, so their gluing in Eq.~\eqref{eq:gluing} implies $T_{\rm Sug}=\bar T_{\rm Sug}$ at the boundary. The coset stress tensor is $T_{\rm cos}=T_{(k)}+T_{(1)}-T_{\rm Sug}$. Subtracting these equalities therefore gives $T_{\rm cos}=\bar T_{\rm cos}$. Since one WZW factor has level one, the coset is the Virasoro minimal model $\mathcal M(k+2,k+3)$, whose chiral algebra is generated by $T_{\rm cos}$~\cite{GKO1986,Quella:2006de}. The boundary therefore preserves
\begin{equation}
  \mathcal A=\frac{\suk{k}\otimes\suk{1}}{\suk{k+1}}\otimes\suk{k+1}.
  \label{eq:algebra}
\end{equation}

We can now enumerate the elementary boundaries that preserve $\mathcal A$. The boundary states of Quella, Runkel, and Watts (QRW)~\cite{QuellaSchomerus2002,Quella:2006de,Ishikawa2002,IshikawaTani2006} carry three representation labels $(\rho_1,\rho_2;\rho)$ for the UV, IR, and diagonal affine algebras, respectively. Each label is twice the spin, so
\begin{equation}
  0\le\rho_1\le k,\quad 0\le\rho_2\le1,\quad 0\le\rho\le k+1.
  \label{eq:labels}
\end{equation}
Let $\ket{\rho_1}$ and $\ket{\rho_2}$ be the Cardy boundary states of the two WZW factors, and let $X^{(N)}_\ell$ denote the Verlinde line labeled by $\ell$ at level $N$. Acting on the product boundary with the line of the diagonal algebra gives the QRW state
\begin{equation}
  \ket{B_{\rho_1,\rho_2;\rho}}=X^{(k+1)}_\rho\,\big(\ket{\rho_1}\otimes\ket{\rho_2}\big).
  \label{eq:line_fusion}
\end{equation}
The values $\rho=0,k+1$ give totally reflecting interfaces, while other values allow energy transmission. The identification $(\rho_1,\rho_2;\rho)\sim(k-\rho_1,1-\rho_2;k+1-\rho)$ gives each state a unique representative with $\rho_2=0$.

These states exhaust the elementary possibilities. As shown in App.~\ref{app:annulus}, their normalized coefficients in the Ishibashi basis form a unitary matrix. Any consistent $\mathcal A$-preserving boundary state can therefore be expanded in QRW states, with coefficients equal to the vacuum multiplicities of its annuli with the QRW boundaries, which are nonnegative integers. For an elementary boundary the vacuum multiplicity with itself, the sum of their squares, is one, so precisely one QRW state occurs.

It remains to impose invariance under the simultaneous center transformation $g\mapsto-g$ on both sides. Its generators in the two factors are the lines $X^{(k)}_k$ and $X^{(1)}_1$~\cite{Chang2019,KormosRunkelWatts2009}, so invariance means $(X^{(k)}_k\otimes X^{(1)}_1)\ket B=\ket B$. At level $N$ the center sends a Cardy label $\ell$ to $N-\ell$. The simultaneous center action preserves each diagonal affine sector and thus commutes with $X^{(k+1)}_\rho$; acting on the Cardy states in Eq.~\eqref{eq:line_fusion} gives~\cite{Quella:2006de}
\[
  (\rho_1,0;\rho)\xrightarrow{\mathbb Z_2}(k-\rho_1,1;\rho)\sim(\rho_1,0;k+1-\rho),
\]
where $\sim$ denotes the label identification above. The state is therefore invariant precisely when $\rho=k+1-\rho$, fixing the diagonal label to
\begin{equation}
  \rho_\ast=\frac{k+1}{2}.
  \label{eq:rho_star}
\end{equation}
This is an integer precisely for odd $k$, in agreement with anomaly matching for the conjectured flow~\cite{LSM1961,AffleckLieb1986,FuruyaOshikawa2017,YaoHsiehOshikawa2019}.

The remaining label $a=\rho_1$ specifies the UV topological line fused (denoted $\star$) onto the undressed interface, so every elementary interface preserving both symmetries takes the form
\begin{equation}
  \ket B=\ket{B_{a,0;\rho_\ast}}=X^{(k)}_a\star\ket{B_{0,0;\rho_\ast}},\quad a=0,\ldots,k.
  \label{eq:theorem}
\end{equation}
Equation~\eqref{eq:theorem} thus identifies the RG interface up to topological dressing; which dressing is realized requires dynamical input.

\textit{Dressing-independent predictions.}---All states in Eq.~\eqref{eq:theorem} have $\rho=\rho_\ast$ and hence the same energy-transmission coefficient~\cite{Quella:2006de},
\begin{equation}
  c_{LR}=\frac{6k}{(k+1)(k+5)}\;\frac{4\cos^2\frac{\pi}{k+3}}{4\cos^2\frac{\pi}{k+3}-1}.
  \label{eq:cLR_main}
\end{equation}
For $k=3$ this gives $c_{LR}=27/32$, the maximum over the QRW family (App.~\ref{app:transmission}).

\begin{table}[t]
\centering
\begin{ruledtabular}
\begin{tabular}{cccl cc}
$k$ & $\rho_\ast$ & $h_1$ & $h_m/h_1$ & $g/g_{0\otimes0}$ & $c_{LR}$\\
\hline
3 & 2 & $1/3$ & $\mathbf{0,1,3},7,12,19$ & $2$ & $27/32$ \\
5 & 3 & $1/4$ & $\mathbf{0,1,3,6},11,17$ & $2.613$ & $1/\sqrt2$ \\
7 & 4 & $1/5$ & $\mathbf{0,1,3,6,10},16$ & $3.236$ & $0.605$ \\
9 & 5 & $1/6$ & $\mathbf{0,1,3,6,10,15}$ & $3.864$ & $0.527$ \\
\end{tabular}
\end{ruledtabular}
\caption{Predictions for the undressed representative $\ket{B_{0,0;\rho_\ast}}$ for $\suk{k}\to\suk{1}$: boundary label, $m=1$ gap $h_1=2/(k+3)$, normalized gaps $h_m/h_1$, $m=0,1,\dots$ [Eq.~\eqref{eq:triangular}; boldface: the Casimir regime $m\le\rho_\ast$], $g$-factor relative to the factorizing boundary [Eq.~\eqref{eq:proposal}], and transmission coefficient [Eq.~\eqref{eq:cLR_main}], which respects $c_{LR}\le1$~\cite{Karch2024}. All dressings have the same $h_m$ and $c_{LR}$. The $g$-factors shown are for $a=0$.}
\label{tab:general}
\end{table}

The lowest scaling dimension in each magnetization sector is also independent of the dressing. On the ring of Fig.~\ref{fig:geometry}, both junctions fold to the same boundary condition $B$. The lattice gaps thus probe states on a strip with $B$ at both ends. When both segments have the same propagation time $\tau$ (length divided by velocity), a strip state of scaling dimension $h$ has excitation energy $\pi h/\tau$. Let $\mathcal H_{BB}$ be the Hilbert space of these states, $L_0$ the Virasoro generator with eigenvalues $h$, and $c_A=c_{\rm UV}+c_{\rm IR}$ the folded central charge. At inverse temperature $\beta$, the open-channel partition function counts the multiplicity $n_h$ of each level in powers of $q=e^{-\pi\beta/\tau}$,
\begin{equation}
  Z_{BB}(q)=\operatorname{Tr}_{\mathcal H_{BB}}q^{L_0-c_A/24}=\sum_h n_h\,q^{h-c_A/24}.
  \label{eq:strip_spectrum}
\end{equation}

The preserved algebra in Eq.~\eqref{eq:algebra} organizes these states into products of coset and diagonal current-algebra representations. A character counts the states in one representation according to their $L_0$ eigenvalues. For $k=3$, let $\chi^{\rm cos}_{(r,s)}$ denote the character of the $\mathcal M(5,6)$ primary with Kac labels $(r,s)$, and $\chi_\ell$ the $\suk{4}$ character of spin $\ell/2$. Suppressing the common argument $q$, we obtain for the undressed boundary $B=B_{0,0;2}$ (App.~\ref{app:annulus})
\begin{align}
  Z_{BB}&=\chi^{\rm cos}_{(1,1)}\big(\chi_0+\chi_2+\chi_4\big)
  +\chi^{\rm cos}_{(1,3)}\big(\chi_0+3\chi_2+\chi_4\big)
  \nonumber\\&\quad
  +\chi^{\rm cos}_{(1,5)}\big(\chi_0+\chi_2+\chi_4\big).
  \label{eq:ZBB_k3}
\end{align}
The coset primaries have weights $0,2/3,3$. In the coset vacuum sector, the affine primaries of spins $0,1,2$ have weights $0,1/3,1$, respectively. The term $\chi^{\rm cos}_{(1,3)}\chi_0$ gives the lowest excited singlet at $h=2/3$.

For every odd $k\ge3$ and every dressing, the coset vacuum is paired with all integrable integer-spin modules of $\widehat{\mathfrak{su}}(2)_N$, where $N=k+1$ (App.~\ref{app:annulus}). Other sectors pair these modules with positive coset weights, so the lowest weight $h_m$ at $J^3_0=S^z=m$ is independent of dressing. By $\mathrm{SU}(2)$ symmetry, $h_{-m}=h_m$. For $0\le m\le\rho_\ast$, the spin-$m$ primary gives
\begin{equation}
  h_m=\frac{m(m+1)}{N+2},\qquad\frac{h_m}{h_1}=\binom{m+1}{2}\quad(m\le\rho_\ast);
  \label{eq:casimir}
\end{equation}
no state is lighter, since the Sugawara form bounds $L_0$ from below by the zero-mode Casimir. For $m>\rho_\ast$ the lowest state is a descendant, and the decomposition of $\suk{N}$ into $U(1)_{2N}$ and $\mathbb Z_N$-parafermion sectors~\cite{FateevZamolodchikov1985,GepnerQiu1987} gives, for all $m$ (App.~\ref{app:weights}),
\begin{equation}
  h_m=\frac{m^2}{N}+\frac{2\,\bar m\,(\rho_\ast-\bar m)}{N(N+2)},\qquad \bar m=m\ {\rm mod}\ \rho_\ast ,
  \label{eq:delta_closed}
\end{equation}
a $U(1)$ charging energy plus a parafermionic twist. The ratios
\begin{equation}
  \frac{h_m}{h_1}=\frac{m^2(N+2)+2\bar m(\rho_\ast-\bar m)}{2N}
  \label{eq:triangular}
\end{equation}
are integers; for $k=3$,
\begin{equation}
  h_m=\frac{m^2}{4}+\frac{1}{12}\,\delta_{m\,\rm odd},\qquad \frac{h_m}{h_1}=0,1,3,7,12,19,\dots
  \label{eq:hm_k3}
\end{equation}
At $m=\rho_\ast+1$, the ratio exceeds the triangular number by one (Table~\ref{tab:general}).

These ratios test the value of $\rho$ selected by center symmetry. Among the undressed representatives $B_{0,0;\rho}$ at $k=3$, only the state with $\rho=2$ contains a spin-two primary in the coset vacuum, giving $h_2/h_1=3$ (\mbox{Table~\ref{tab:results}}). For $\rho\ne\rho_\ast$, non-invertible dressing can change these ratios. Since the microscopic junction breaks one-site translation, the comparison also tests center symmetry of its continuum limit (App.~\ref{app:numerics}).

\begin{figure*}[t]
\centering
\includegraphics[width=0.9\textwidth]{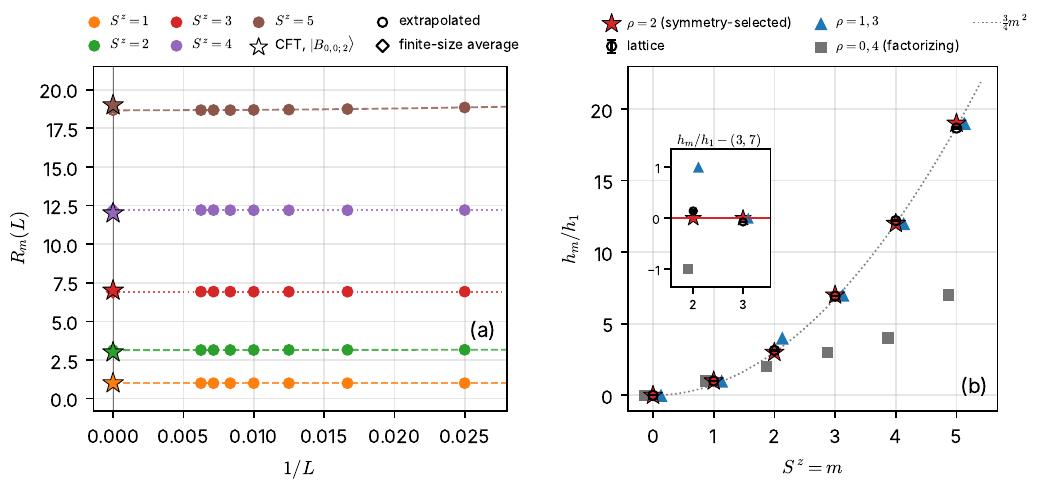}
\caption{Defect spectrum of the $\suk{3}\to\suk{1}$ RG interface, $L=40$--$160$. (a) Normalized gaps~\eqref{eq:ratio} versus $1/L$ ($S^z=1$ is one by construction). Dashed lines and open circles ($S^z=2,5$): fits $c+a/L^{b}$ with $b$ free and the extrapolated values; dotted lines and open diamonds ($S^z=3,4$): finite-size averages with half-ranges; stars: Eq.~\eqref{eq:hm_k3}. (b) The same estimates compared with the undressed representatives $B_{0,0;\rho}$ (non-invertible dressing can change these values away from $\rho=\rho_\ast$): $\rho=2$, $\rho=1,3$, and the factorizing $\rho=0,4$, offset horizontally; dotted: $U(1)_8$ charging energy $\tfrac34m^2$. Inset: deviation from $(3,7)$ in the sectors $S^z=2,3$.}
\label{fig:defect_spectrum}
\end{figure*}

\textit{Distinguishing the dressings.}---At $\rho=\rho_\ast$ the lowest $S^z=0$ level is the vacuum, $h_0=0$, for every dressing. The first excited level in this sector distinguishes the candidates. The undressed boundary ($a=0$) and its invertible dressing ($a=k$) have identical strip spectra, with all weights integer multiples of the first magnetization gap (App.~\ref{app:lowlying}),
\begin{equation}
  h\in h_1\,\mathbb Z_{\ge0},\qquad h_1=\frac{2}{N+2}.
  \label{eq:tower}
\end{equation}
For $k=3$, the first excited $S^z=0$ level of the $a=0,3$ pair belongs to the lowest triplet, at $h/h_1=1$ [Eq.~\eqref{eq:ZBB_k3}]. For $a=1,2$ the first excitation in this sector is instead the singlet $\phi_{3,3}$ in the diagonal vacuum sector, at $h/h_1=1/5$ [Eq.~\eqref{eq:ZB}].

Among the boundaries in Eq.~\eqref{eq:theorem}, the pair $a=0,k$ also minimizes the interface $g$-factor~\cite{AffleckLudwig1991}. Fusing the UV line $X^{(k)}_a$ onto the undressed boundary multiplies $g$ by its quantum dimension $d_a$, which is one for $a=0,k$ and greater than one for $0<a<k$ (App.~\ref{app:dressing}). Let $g_{0\otimes0}$ denote the $g$-factor of the product of the two vacuum Cardy boundaries and $S^{(k+1)}$ the modular $S$-matrix of the diagonal current algebra. Then
\begin{equation}
  \frac{g_{B_{a,0;\rho_\ast}}}{g_{0\otimes0}}=d_a\,\frac{S^{(k+1)}_{\rho_\ast0}}{S^{(k+1)}_{00}}=\frac{d_a}{\sin\frac{\pi}{k+3}}.
  \label{eq:proposal}
\end{equation}
For this pair, the only nonidentity relevant $\mathrm{SU}(2)$ singlet, $\phi_{1,3}$, is center-odd, and there is no marginal singlet. The pair is therefore stable under perturbations preserving both symmetries (App.~\ref{app:annulus}).

\textit{Lattice test.}---We use the spin-$3/2$ chain~\cite{Michaud2012,Michaud:2013nq,chepiga2020floating}
\begin{equation}
\begin{aligned}
H = \sum_{i} \Big[ & J_1 \vec S_i\!\cdot\!\vec S_{i+1} + J_2 \vec S_i\!\cdot\!\vec S_{i+2}
\\
&+ J_{3,i}\big((\vec S_{i-1}\!\cdot\!\vec S_i)(\vec S_i\!\cdot\!\vec S_{i+1}) + \text{H.c.}\big)\Big]
\end{aligned}
\label{eq:HWZW}
\end{equation}
on a periodic ring of $L$ sites, with $J_1=1$, $J_2=0.1$, and, labeling each three-site term by its central site $i$, $J_{3,i}=J_3^c=0.05128$ on the right half, the $\suk{3}$ critical point of this model~\cite{Michaud:2013nq,chepiga2020floating}, and $J_{3,i}=0$ on the left half, in the $\suk{1}$ phase; the ring is symmetric under the reflection exchanging the two junctions. Homogeneous rings at these two couplings reproduce the $\suk{1}$ and $\suk{3}$ spectra within about $3\%$ and $2\%$ (App.~\ref{app:homogeneous}). We compute the lowest energy $E(S^z)$ in each sector $S^z=0,\dots,5$ for $L=40$--$160$ by density-matrix renormalization group (DMRG) with bond dimensions up to $4000$ and a truncation error $10^{-8}$, and form
\begin{equation}
  R_m(L)=\frac{E(S^z\!=\!m)-E(0)}{E(1)-E(0)} .
  \label{eq:ratio}
\end{equation}
At the fixed point, and for equal propagation times on the two sides, $R_m(L)\to h_m/h_1$; the segment-length ratio can compensate for unequal bulk velocities by tuning $\tau_{\rm IR}/\tau_{\rm UV}=(L_{\rm IR}/L_{\rm UV})(v_{\rm UV}/v_{\rm IR})$ towards unity (App.~\ref{app:numerics}).

\begin{table}[t]
\centering
\begin{ruledtabular}
\begin{tabular}{lcccc}
$S^z$ & 2 & 3 & 4 & 5 \\
\hline
lattice & $3.139(9)$ & $6.925(2)$ & $12.199(8)$ & $18.66(9)$ \\
$\rho=2$ (symmetry) & 3 & 7 & 12 & 19 \\
$\rho=1,3$ & 4 & 7 & 12 & 19 \\
$\rho=0,4$ (factorizing) & 2 & 3 & 4 & 7 \\
\end{tabular}
\end{ruledtabular}
\caption{Normalized gaps $h_m/h_1$ for $\suk{3}\to\suk{1}$ compared with the undressed representatives $B_{0,0;\rho}$ ($h_1=1$ at $\rho=0,4$, else $1/3$); non-invertible dressing can change these values except at $\rho=2$. Lattice values are extrapolations ($S^z=2,5$) or finite-size averages ($S^z=3,4$); parentheses are stability measures only (App.~\ref{app:numerics}).}
\label{tab:results}
\end{table}

Figure~\ref{fig:defect_spectrum} and Table~\ref{tab:results} show the result; the ratios drift only slowly with $L$. We extrapolate $S^z=2,5$ with the ansatz $c+a/L^b$ and quote finite-size averages for $S^z=3,4$, which vary by less than $0.2\%$ (App.~\ref{app:numerics}). The estimates agree with Eq.~\eqref{eq:hm_k3} to within $2\%$ for $S^z=3,4,5$ and $5\%$ for $S^z=2$, and the same pattern is obtained at $J_2=0$, $J_3=0.063$ (App.~\ref{app:numerics}). The sectors $m=2,3$ discriminate among the representatives of Table~\ref{tab:results}, $(3,7)$ for $\rho=2$ against $(4,7)$ and $(2,3)$, and the data $(3.14,6.93)$ are closest to the first. With $L_{\rm IR}=2L_{\rm UV}$, $R_2$ decreases from $3.101$ to $3.046$ over $L=24$--$72$ and extrapolates to $3.01(3)$, which identifies the propagation-time imbalance as the main source of the remaining deviation; the other sectors stay within a few percent (App.~\ref{app:numerics}). Non-invertibly dressed states with other $\rho$ are addressed by the excited level discussed next.

The first excited state of the $S^z=0$ sector tests the dressing. On the $L=40$ ring its gap coincides with the $S^z=1$ gap within the numerical precision,
\begin{equation}
  E_2(S^z\!=\!0)-E(0)\simeq E(1)-E(0),
  \label{eq:dressing_test}
\end{equation}
i.e.\ the normalized level is $1$, as for $B_{0,0;2}$, whereas $B_{1,0;2}$ and $B_{2,0;2}$ predict $0.2$ (App.~\ref{app:numerics}). This is the $S^z=0$ component of the lowest triplet, as expected for the undressed boundary; the invertible dressing $B_{3,0;2}$ has the same spectrum and is not distinguished. The test reaches further: by the fusion rules every $B_{a,0;\rho}$ with $0<a<k$, $1\le\rho\le k$ contains the singlet $\phi_{3,3}\otimes\chi_0$ at $h/h_1=1/(k+2)$ (App.~\ref{app:annulus}), so the absence of a singlet below the first triplet distinguishes the minimal-$g$ pair from all non-invertibly dressed transmitting representatives; dressed factorizing boundaries ($\rho=0,k+1$) are not excluded. For the spin-$5/2$ chain, which tests the proposed $\suk{5}\to\suk{1}$ flow, rings with $L=20$--$40$ give $3.04$, $6.24$, $10.65$, $16.1$ for $S^z=2,\dots,5$, within a few percent of the pattern $3,6,11,17$ of $\rho_\ast=3$; the discriminating sector $S^z=3$ lies $4\%$ from $6$ and $11\%$ from the $7$ of $\rho=2,4$, and $S^z=2$ disfavors the undressed representatives with $\rho=0,1,5,6$. The extra singlet of the dressed candidates, here at $h/h_1=1/7$, is again absent: on the $L=20$ ring the first excited $S^z=0$ level lies at the $S^z=1$ gap (App.~\ref{app:su25}).

\textit{Discussion.}---Global $\mathrm{SU}(2)$ symmetry and the $\mathbb Z_2$ center symmetry together fix the elementary WZW RG interface up to topological dressing. Although the $k=3$ and $k=5$ flows are expected to reach the same IR fixed point, this identification predicts a distinct pattern of gap ratios across magnetization sectors for each flow. The spin-chain spectra reproduce both patterns, providing evidence for the conjectured flows. A natural question is whether symmetry also fixes the RG interfaces of the higher-rank $SU(N)_k\to SU(N)_1$ flows~\cite{Lecheminant2015,FukusumiNakashiba2026}. Here the coset algebra contains additional currents, and one must determine which of them the interface preserves.

\begin{acknowledgments}
\textit{Acknowledgement.}---YK is supported by the INAMORI Frontier Program at Kyushu University, JSPS KAKENHI Grant Number 23K20046, MEXT Supporting Pioneering Research through AI for 1,000 Discovery Challenges (SPReAD) Grant Number JPMXP1726262823, and JST PRESTO Grant Number JPMJPR26A6.
TO is supported by JST SPRING, Grant Number JPMJSP2110.
We used \mbox{ChatGPT} (OpenAI, GPT-6.2 Astra) and Claude (Anthropic, Opus 5.5 and Fable 5.2) to assist with drafting, writing numerical codes and checking the manuscript under our direction. The authors are responsible for all content and conclusions.
\end{acknowledgments}

\bibliographystyle{apsrev4-1}
\bibliography{main-defect}

@article{Oshikawa1996,
  author={Oshikawa, Masaki and Affleck, Ian}, title={Defect lines in the {Ising} model and boundary states on orbifolds},
  journal={Phys. Rev. Lett.}, volume={77}, pages={2604}, year={1996}, eprint={hep-th/9606177}}

@article{Oshikawa1997,
  author={Oshikawa, Masaki and Affleck, Ian}, title={Boundary conformal field theory approach to the critical two-dimensional {Ising} model with a defect line},
  journal={Nucl. Phys. B}, volume={495}, pages={533}, year={1997}, eprint={cond-mat/9612187}}

@article{Bachas2002,
  author={Bachas, Constantin and de Boer, Jan and Dijkgraaf, Robbert and Ooguri, Hirosi}, title={Permeable conformal walls and holography},
  journal={JHEP}, volume={06}, pages={027}, year={2002}, eprint={hep-th/0111210}}

@article{Wong1993,
  author={Wong, Eugene and Affleck, Ian}, title={Tunneling in quantum wires: a boundary conformal field theory approach},
  journal={Nucl. Phys. B}, volume={417}, pages={403}, year={1994}, eprint={cond-mat/9311040}}

@article{PetkovaZuber2001,
  author={Petkova, V. B. and Zuber, J.-B.}, title={Generalised twisted partition functions},
  journal={Phys. Lett. B}, volume={504}, pages={157}, year={2001}, eprint={hep-th/0011021}}

@article{FFRS2004,
  author={Fr{\"o}hlich, J{\"u}rg and Fuchs, J{\"u}rgen and Runkel, Ingo and Schweigert, Christoph}, title={{Kramers-Wannier} duality from conformal defects},
  journal={Phys. Rev. Lett.}, volume={93}, pages={070601}, year={2004}, eprint={cond-mat/0404051}}

@article{FFRS2007,
  author={Fr{\"o}hlich, J{\"u}rg and Fuchs, J{\"u}rgen and Runkel, Ingo and Schweigert, Christoph}, title={Duality and defects in rational conformal field theory},
  journal={Nucl. Phys. B}, volume={763}, pages={354}, year={2007}, eprint={hep-th/0607247}}

@article{BachasBrunner2008,
  author={Bachas, Constantin and Brunner, Ilka}, title={Fusion of conformal interfaces},
  journal={JHEP}, volume={02}, pages={085}, year={2008}, eprint={0712.0076}}

@article{Cardy1989,
  author={Cardy, John L.}, title={Boundary conditions, fusion rules and the {Verlinde} formula},
  journal={Nucl. Phys. B}, volume={324}, pages={581}, year={1989}}

@article{Ishibashi1989,
  author={Ishibashi, Nobuyuki}, title={The boundary and crosscap states in conformal field theories},
  journal={Mod. Phys. Lett. A}, volume={4}, pages={251}, year={1989}}

@article{Verlinde1988,
  author={Verlinde, Erik}, title={Fusion rules and modular transformations in {2D} conformal field theory},
  journal={Nucl. Phys. B}, volume={300}, pages={360}, year={1988}}

@article{GKO1986,
  author={Goddard, P. and Kent, A. and Olive, D.}, title={Unitary representations of the {Virasoro} and super-{Virasoro} algebras},
  journal={Commun. Math. Phys.}, volume={103}, pages={105}, year={1986}}

@book{DiFrancesco1997,
  author={Di Francesco, Philippe and Mathieu, Pierre and S{\'e}n{\'e}chal, David}, title={Conformal Field Theory},
  publisher={Springer}, address={New York}, year={1997}}

@book{RecknagelSchomerus2013,
  author={Recknagel, Andreas and Schomerus, Volker}, title={Boundary Conformal Field Theory and the Worldsheet Approach to {D}-Branes},
  publisher={Cambridge University Press}, year={2013}}

@article{AffleckLudwig1991,
  author={Affleck, Ian and Ludwig, Andreas W. W.}, title={Universal noninteger ``ground-state degeneracy'' in critical quantum systems},
  journal={Phys. Rev. Lett.}, volume={67}, pages={161}, year={1991}}

@article{FriedanKonechny2004,
  author={Friedan, Daniel and Konechny, Anatoly}, title={On the boundary entropy of one-dimensional quantum systems at low temperature},
  journal={Phys. Rev. Lett.}, volume={93}, pages={030402}, year={2004}, eprint={hep-th/0312197}}

@article{Casini2016g,
  author={Casini, Horacio and Salazar Landea, Ignacio and Torroba, Gonzalo}, title={The g-theorem and quantum information theory},
  journal={JHEP}, volume={10}, pages={140}, year={2016}, eprint={1607.00390}}

@article{Gaiotto:2012np,
  author={Gaiotto, Davide}, title={Domain walls for two-dimensional renormalization group flows},
  journal={JHEP}, volume={12}, pages={103}, year={2012}, eprint={1201.0767}}

@article{BrunnerRoggenkamp2008,
  author={Brunner, Ilka and Roggenkamp, Daniel}, title={Defects and bulk perturbations of boundary {Landau-Ginzburg} orbifolds},
  journal={JHEP}, volume={04}, pages={001}, year={2008}, eprint={0712.0188}}

@article{Poghosyan2013mixing,
  author={Poghosyan, Armen and Poghosyan, Hayk}, title={Mixing with descendant fields in perturbed minimal {CFT} models},
  journal={JHEP}, volume={10}, pages={131}, year={2013}, eprint={1305.6066}}

@article{Poghossian2014,
  author={Poghossian, Rubik}, title={Two dimensional renormalization group flows in next to leading order},
  journal={JHEP}, volume={01}, pages={167}, year={2014}, eprint={1303.3015}}

@article{Poghosyan2014,
  author={Poghosyan, Gabriel and Poghosyan, Hasmik}, title={{RG} domain wall for the {$N=1$} minimal superconformal models},
  journal={JHEP}, volume={05}, pages={043}, year={2015}, eprint={1412.6710}}

@article{Poghosyan2022W3,
  author={Poghosyan, Hasmik and Poghossian, Rubik}, title={{RG} flow between {$W_3$} minimal models by perturbation and domain wall approaches},
  journal={JHEP}, volume={08}, pages={307}, year={2022}, eprint={2205.05091}}

@article{Poghosyan2023A2,
  author={Poghosyan, Armen and Poghosyan, Hasmik}, title={A note on {RG} domain wall between successive {$A_2^{(p)}$} minimal models},
  journal={JHEP}, volume={08}, pages={072}, year={2023}, eprint={2305.05997}}

@article{Stanishkov2016a,
  author={Stanishkov, Marian}, title={{RG} domain wall for the general {$\widehat{su}(2)$} coset models},
  journal={JHEP}, volume={08}, pages={096}, year={2016}, eprint={1606.03605}}

@article{Stanishkov2016b,
  author={Stanishkov, Marian}, title={Second order {RG} flow in general {$\widehat{su}(2)$} coset models},
  journal={JHEP}, volume={09}, pages={040}, year={2016}, eprint={1606.04328}}

@article{AhnStanishkov2014,
  author={Ahn, Changrim and Stanishkov, Marian}, title={On the renormalization group flow in two dimensional superconformal models},
  journal={Nucl. Phys. B}, volume={885}, pages={713}, year={2014}, eprint={1404.7628}}

@article{KonechnySchmidtColinet2014,
  author={Konechny, Anatoly and Schmidt-Colinet, Cornelius}, title={Entropy of conformal perturbation defects},
  journal={J. Phys. A}, volume={47}, pages={485401}, year={2014}, eprint={1407.6444}}

@article{Cogburn2024,
  author={Cogburn, Cameron V. and Fitzpatrick, A. Liam and Geng, Hao}, title={{CFT} and lattice correlators near an {RG} domain wall between minimal models},
  journal={SciPost Phys. Core}, volume={7}, pages={021}, year={2024}, eprint={2308.00737}}

@article{AntunesChenRong2026,
  author={Antunes, Ant{\'o}nio and Chen, Jin and Rong, Junchen},
  title={Rational conformal interfaces between the {Ising} and tricritical {Ising} models from a tetrality web},
  year={2026}, eprint={2611.18831}, archivePrefix={arXiv}, primaryClass={hep-th}}

@article{AntunesRong:2026w3isingtim,
    author = "Antunes, Ant{\'o}nio and Rong, Junchen",
    title = "{On the Ising/tricritical-Ising interface: From ${W}_3$ symmetry to Rydberg atoms}",
    eprint = "2605.23848",
    archivePrefix = "arXiv",
    primaryClass = "hep-th",
    month = "5",
    year = "2026"
}

@article{Zhang2026,
  author={Zhang, Jin-Rui and Jin, Jing-Hao and Chen, Ting-Kai and Chen, Jin}, title={Defect conformal manifolds along {RG} domain walls between {$\mathbb{Z}_N$}-parafermions and minimal models},
  journal={JHEP}, volume={09}, pages={123}, year={2026}, eprint={2605.24978}}

@article{KormosRunkelWatts2009,
  author={Kormos, M{\'a}rton and Runkel, Ingo and Watts, Gerard M. T.}, title={Defect flows in minimal models},
  journal={JHEP}, volume={11}, pages={057}, year={2009}, eprint={0907.1497}}

@article{Chang2019,
  author={Chang, Chi-Ming and Lin, Ying-Hsuan and Shao, Shu-Heng and Wang, Yifan and Yin, Xi}, title={Topological defect lines and renormalization group flows in two dimensions},
  journal={JHEP}, volume={01}, pages={026}, year={2019}, eprint={1802.04445}}

@article{TanakaNakayama2024,
  author={Nakayama, Yu and Tanaka, Takahiro}, title={Infinitely many new renormalization group flows between {Virasoro} minimal models from non-invertible symmetries},
  journal={JHEP}, volume={11}, pages={137}, year={2024}, eprint={2407.21353}}

@article{AmbrosinoNegro2025,
  author={Ambrosino, Federico and Negro, Stefano}, title={Minimal model renormalization group flows: noninvertible symmetries and nonperturbative description},
  journal={Phys. Rev. Lett.}, volume={135}, pages={021602}, year={2025}, eprint={2501.07511}}

@article{FuruyaOshikawa2017,
  author={Furuya, Shunsuke C. and Oshikawa, Masaki}, title={Symmetry protection of critical phases and a global anomaly in $1+1$ dimensions},
  journal={Phys. Rev. Lett.}, volume={118}, pages={021601}, year={2017}, eprint={1503.07292}, doi={10.1103/PhysRevLett.118.021601}}

@article{YaoHsiehOshikawa2019,
  author={Yao, Yuan and Hsieh, Chang-Tse and Oshikawa, Masaki}, title={Anomaly matching and symmetry-protected critical phases in {$SU(N)$} spin systems in $1+1$ dimensions},
  journal={Phys. Rev. Lett.}, volume={123}, pages={180201}, year={2019}, eprint={1805.06885}, doi={10.1103/PhysRevLett.123.180201}}

@article{Quella:2006de,
  author={Quella, Thomas and Runkel, Ingo and Watts, Gerard M. T.}, title={Reflection and transmission for conformal defects},
  journal={JHEP}, volume={04}, pages={095}, year={2007}, eprint={hep-th/0611296}}

@article{Meineri2020,
  author={Meineri, Marco and Penedones, Jo{\~a}o and Rousset, Antonin}, title={Colliders and conformal interfaces},
  journal={JHEP}, volume={02}, pages={138}, year={2020}, eprint={1904.10974}}

@article{Karch2024,
  author={Karch, Andreas and Kusuki, Yuya and Ooguri, Hirosi and Sun, Hao-Yu and Wang, Mianqi}, title={Universal bound on effective central charge and its saturation},
  journal={Phys. Rev. Lett.}, volume={133}, pages={091604}, year={2024}, eprint={2404.01515}}

@article{Furuta2025,
  author={Furuta, Yuma and Kusuki, Yuya and Onagi, Toshiki}, title={Transmission coefficients from phantom currents},
  journal={Phys. Rev. D}, volume={113}, pages={045008}, year={2026}, eprint={2511.00356}}

@article{Haldane1983,
  author={Haldane, F. D. M.}, title={Nonlinear field theory of large-spin {Heisenberg} antiferromagnets: Semiclassically quantized solitons of the one-dimensional easy-axis {N\'eel} state},
  journal={Phys. Rev. Lett.}, volume={50}, pages={1153}, year={1983}}

@article{Affleck1985,
  author={Affleck, Ian}, title={Critical behavior of two-dimensional systems with continuous symmetries},
  journal={Phys. Rev. Lett.}, volume={55}, pages={1355}, year={1985}}

@article{Affleck1986NPB,
  author={Affleck, Ian}, title={Exact critical exponents for quantum spin chains, non-linear $\sigma$-models at $\theta=\pi$ and the quantum {Hall} effect},
  journal={Nucl. Phys. B}, volume={265}, pages={409}, year={1986}}

@article{Affleck1987,
  author={Affleck, Ian and Haldane, F. D. M.}, title={Critical theory of quantum spin chains},
  journal={Phys. Rev. B}, volume={36}, pages={5291}, year={1987}}

@article{Konik2015,
  author={Konik, R. M. and P{\'a}lmai, T. and Tak{\'a}cs, G. and Tsvelik, A. M.}, title={Studying the perturbed {Wess-Zumino-Novikov-Witten} {SU(2)$_k$} theory using the truncated conformal spectrum approach},
  journal={Nucl. Phys. B}, volume={899}, pages={547}, year={2015}, eprint={1505.03860}}

@article{James2018,
  author={James, A. J. A. and Konik, R. M. and Lecheminant, P. and Robinson, N. J. and Tsvelik, A. M.}, title={Non-perturbative methodologies for low-dimensional strongly-correlated systems: From non-{Abelian} bosonization to truncated spectrum methods},
  journal={Rept. Prog. Phys.}, volume={81}, pages={046002}, year={2018}, eprint={1703.08421}}

@article{LSM1961,
  author={Lieb, Elliott and Schultz, Theodore and Mattis, Daniel}, title={Two soluble models of an antiferromagnetic chain},
  journal={Ann. Phys.}, volume={16}, pages={407}, year={1961}}

@article{AffleckLieb1986,
  author={Affleck, Ian and Lieb, Elliott H.}, title={A proof of part of {Haldane's} conjecture on spin chains},
  journal={Lett. Math. Phys.}, volume={12}, pages={57}, year={1986}}

@article{Michaud2012,
  author={Michaud, Fr{\'e}d{\'e}ric and Vernay, Fran{\c{c}}ois and Manmana, Salvatore R. and Mila, Fr{\'e}d{\'e}ric}, title={Antiferromagnetic spin-{$S$} chains with exactly dimerized ground states},
  journal={Phys. Rev. Lett.}, volume={108}, pages={127202}, year={2012}, eprint={1110.3394}}

@article{Michaud:2013nq,
  author={Michaud, Fr{\'e}d{\'e}ric and Manmana, Salvatore R. and Mila, Fr{\'e}d{\'e}ric}, title={Realization of higher {Wess-Zumino-Witten} models in spin chains},
  journal={Phys. Rev. B}, volume={87}, pages={140404(R)}, year={2013}, eprint={1301.5719}}

@article{chepiga2020floating,
  author={Chepiga, Natalia and Affleck, Ian and Mila, Fr{\'e}d{\'e}ric}, title={Floating, critical, and dimerized phases in a frustrated spin-3/2 chain},
  journal={Phys. Rev. B}, volume={101}, pages={174407}, year={2020}, eprint={2002.08982}}

@article{Chepiga:2022ciu,
  author={Chepiga, Natalia and Affleck, Ian and Mila, Fr{\'e}d{\'e}ric}, title={From {SU(2)$_5$} to {SU(2)$_3$} {Wess-Zumino-Witten} transitions in a frustrated spin-5/2 chain},
  journal={Phys. Rev. B}, volume={105}, pages={174402}, year={2022}, eprint={2202.05087}}

@article{Cardy1986log,
  author={Cardy, John L.}, title={Logarithmic corrections to finite-size scaling in strips},
  journal={J. Phys. A}, volume={19}, pages={L1093}, year={1986}}

@article{AGSZ1989,
  author={Affleck, I. and Gepner, D. and Schulz, H. J. and Ziman, T.}, title={Critical behaviour of spin-$s$ {Heisenberg} antiferromagnetic chains: analytic and numerical results},
  journal={J. Phys. A}, volume={22}, pages={511}, year={1989}}

@article{FateevZamolodchikov1985,
  author={Zamolodchikov, A. B. and Fateev, V. A.}, title={Nonlocal (parafermion) currents in two-dimensional conformal quantum field theory and self-dual critical points in {$Z_N$}-symmetric statistical systems},
  journal={Sov. Phys. JETP}, volume={62}, pages={215}, year={1985}}

@article{GepnerQiu1987,
  author={Gepner, Doron and Qiu, Zongan}, title={Modular invariant partition functions for parafermionic field theories},
  journal={Nucl. Phys. B}, volume={285}, pages={423}, year={1987}}

@article{Chen2026a,
  author={Chen, Jin}, title={Renormalization group domain walls from phantom symmetries},
  year={2026}, eprint={2610.23529}, archivePrefix={arXiv}, primaryClass={hep-th}}

@article{Antinucci2025,
  author={Antinucci, Andrea and Copetti, Christian and Galati, Giovanni and Rizi, Giovanni}, title={Defect conformal manifolds from phantom noninvertible symmetries},
  journal={Phys. Rev. Lett.}, volume={135}, pages={211602}, year={2025}, eprint={2505.09668}}

@article{BenedettiFendleyMagan2026,
  author={Benedetti, Valentin and Fendley, Paul and Magan, Javier M.}, title={Non-invertible symmetries and selection rules for {RG} flows of coset models},
  year={2026}, eprint={2603.09591}, archivePrefix={arXiv}}

@article{FTZ2021,
  author={Fukusumi, Yoshiki and Tachikawa, Yuji and Zheng, Yunqin}, title={Fermionization and boundary states in $1+1$ dimensions},
  journal={SciPost Phys.}, volume={11}, pages={082}, year={2021}, eprint={2103.00746}}

@article{Ishikawa2002,
  author={Ishikawa, Hiroshi}, title={Boundary states in coset conformal field theories},
  journal={Nucl. Phys. B}, volume={629}, pages={209}, year={2002}, eprint={hep-th/0111230}}

@article{IshikawaTani2006,
  author={Ishikawa, Hiroshi and Tani, Taro}, title={Twisted boundary states and representation of generalized fusion algebra},
  journal={Nucl. Phys. B}, volume={739}, pages={328}, year={2006}, eprint={hep-th/0510242}}

@article{SchellekensWarner1986,
  author={Schellekens, A. N. and Warner, N. P.}, title={Conformal subalgebras of {Kac-Moody} algebras},
  journal={Phys. Rev. D}, volume={34}, pages={3092}, year={1986}}

@article{BaisBouwknegt1987,
  author={Bais, F. A. and Bouwknegt, P.}, title={A classification of subgroup truncations of the bosonic string},
  journal={Nucl. Phys. B}, volume={279}, pages={561}, year={1987}}

@article{ThorngrenWang2021,
  author={Thorngren, Ryan and Wang, Yifan}, title={Anomalous symmetries end at the boundary},
  journal={JHEP}, volume={09}, pages={017}, year={2021}, eprint={2012.15861}}

@article{QuellaSchomerus2002,
  author={Quella, Thomas and Schomerus, Volker}, title={Symmetry breaking boundary states and defect lines},
  journal={JHEP}, volume={06}, pages={028}, year={2002}, eprint={hep-th/0203161}}

@article{NumasawaYamaguchi2018,
  author={Numasawa, Tokiro and Yamaguchi, Satoshi}, title={Mixed global anomalies and boundary conformal field theories},
  journal={JHEP}, volume={11}, pages={202}, year={2018}, eprint={1712.09361}, doi={10.1007/JHEP11(2018)202}}

@article{LiHsiehYaoOshikawa2024,
  author={Li, Linhao and Hsieh, Chang-Tse and Yao, Yuan and Oshikawa, Masaki}, title={Boundary conditions and anomalies of conformal field theories in $1+1$ dimensions},
  journal={Phys. Rev. B}, volume={110}, pages={045118}, year={2024}, eprint={2205.11190}, doi={10.1103/PhysRevB.110.045118}}

@article{JensenShaverinYarom2018,
  author={Jensen, Kristan and Shaverin, Evgeny and Yarom, Amos}, title={'t Hooft anomalies and boundaries},
  journal={JHEP}, volume={01}, pages={085}, year={2018}, eprint={1710.07299}, doi={10.1007/JHEP01(2018)085}}

@article{HanTiwariHsiehRyu2017,
  author={Han, Bo and Tiwari, Apoorv and Hsieh, Chang-Tse and Ryu, Shinsei}, title={Boundary conformal field theory and symmetry-protected topological phases in $2+1$ dimensions},
  journal={Phys. Rev. B}, volume={96}, pages={125105}, year={2017}, eprint={1704.01193}, doi={10.1103/PhysRevB.96.125105}}

@article{ChoiRayhaunSanghaviShao2023,
  author={Choi, Yichul and Rayhaun, Brandon C. and Sanghavi, Yaman and Shao, Shu-Heng}, title={Remarks on boundaries, anomalies, and noninvertible symmetries},
  journal={Phys. Rev. D}, volume={108}, pages={125005}, year={2023}, eprint={2305.09713}, doi={10.1103/PhysRevD.108.125005}}

@article{Lecheminant2015,
  author={Lecheminant, Philippe}, title={Massless renormalization group flow in {$SU(N)_k$} perturbed conformal field theory},
  journal={Nucl. Phys. B}, volume={901}, pages={510}, year={2015}, eprint={1509.01680}, doi={10.1016/j.nuclphysb.2015.11.004}}

@article{FukusumiNakashiba2026,
  author={Fukusumi, Yoshiki and Nakashiba, Shuma}, title={Difference between pair creation and anomaly matching: Nontrivial aspects of invertible symmetries and defects in {$SU(N)$} models},
  year={2026}, eprint={2610.21478}, archivePrefix={arXiv}, primaryClass={hep-th}}

\clearpage
\appendix
\setcounter{secnumdepth}{3}
\onecolumngrid

\section{Open-channel spectrum and Cardy consistency of the QRW states}
\label{app:annulus}
We derive the strip partition function for boundaries $B=B_{\rho_1,\rho_2;\rho}$ and $B'=B_{\rho'_1,\rho'_2;\rho'}$, with labels as in Eq.~\eqref{eq:labels}. Setting $B'=B$ gives the spectrum in Eq.~\eqref{eq:strip_spectrum}. Write $S^{(k)}$, $S^{(1)}$, $S^{(k+1)}$ for the modular $S$-matrices of the three affine algebras; they are real and symmetric, and satisfy $S^{(N)}_{N-a,\,b}=(-1)^{b}S^{(N)}_{a,b}$. The coset $\mathcal M(k+2,k+3)=\suk{k}\otimes\suk{1}/\suk{k+1}$~\cite{GKO1986} has fields $(\mu_1,\mu_2;\mu)$ with $\mu_1+\mu_2+\mu\in2\mathbb Z$, identified under $(\mu_1,\mu_2;\mu)\sim(k-\mu_1,1-\mu_2;k+1-\mu)$, and conformal weights
\begin{equation}
  h^{\rm cos}_{(\mu_1,\mu_2;\mu)}=h_{r,s},\quad r=\mu_1+1,\ s=\mu+1,\quad
  h_{r,s}=\frac{\big[(k+3)r-(k+2)s\big]^2-1}{4(k+2)(k+3)} .
\end{equation}
We denote a coset character by $\chi^{\rm cos}_{(\mu_1,\mu_2;\mu)}$ and the spin-$\ell/2$ character at level $K$ by $\chi^{(K)}_\ell$. The open-channel parameter $q$ and the closed-channel parameter $\tilde q$ are related by the modular $S$ transformation. The coset $S$-matrix on identification classes is $S^{\rm cos}_{[\mu][\nu]}=2\,S^{(k)}_{\mu_1\nu_1}S^{(1)}_{\mu_2\nu_2}S^{(k+1)}_{\mu\nu}$, so that
\begin{equation}
  \chi^{\rm cos}_{(\mu_1,\mu_2;\mu)}(\tilde q)=\sum_{(\nu_1,\nu_2;\nu)}S^{(k)}_{\mu_1\nu_1}S^{(1)}_{\mu_2\nu_2}S^{(k+1)}_{\mu\nu}\,\chi^{\rm cos}_{(\nu_1,\nu_2;\nu)}(q),
  \label{eq:cosetS}
\end{equation}
where the sum runs over all triples satisfying the selection rule (each class is counted twice, which compensates the factor $2$).

The Ishibashi states for the gluing algebra $\mathcal A$ are labeled by the triples $(\mu_1,\mu_2;\mu)$ with $\mu_1+\mu_2+\mu$ even; the triple fixes both the bulk sector $(\mu_1,\mu_2)$ and the $\mathcal A$-representation $(\mu_1,\mu_2;\mu)\otimes\mu$, and two identified triples label different $\mathcal A$-representations unless $\mu=(k+1)/2$, where they label the same representation appearing in the two bulk sectors $(\mu_1,\mu_2)$ and $(k-\mu_1,1-\mu_2)$; the $(k+1)(k+2)$ Ishibashi states thus span $(k+1)(k+2)-(k+1)/2$ distinct representations ($18$ for $k=3$), the pairs at $\mu=(k+1)/2$ providing two orthogonal Ishibashi states for the same representation~\cite{QuellaSchomerus2002}. Their overlaps are $\langle\!\langle(\mu_1,\mu_2;\mu)|\otimes\langle\!\langle\mu|\,\tilde q^{H}\,|(\mu_1',\mu_2';\mu')\rangle\!\rangle\otimes|\mu'\rangle\!\rangle=\delta\,\chi^{\rm cos}_{(\mu_1,\mu_2;\mu)}(\tilde q)\chi^{(k+1)}_{\mu}(\tilde q)$. In this basis the QRW states of Eq.~\eqref{eq:line_fusion} read~\cite{Quella:2006de}
\begin{align}
  \ket{B_{\rho_1,\rho_2;\rho}}
  =\!\!\sum_{\substack{\mu_1,\mu_2,\mu\\ \mu_1+\mu_2+\mu\in2\mathbb Z}}\!\!
  &\frac{S^{(k)}_{\rho_1\mu_1}S^{(1)}_{\rho_2\mu_2}}{\sqrt{S^{(k)}_{0\mu_1}S^{(1)}_{0\mu_2}}}\,
  \frac{S^{(k+1)}_{\rho\mu}}{S^{(k+1)}_{0\mu}}
  \nonumber\\
  &\times\kett{(\mu_1,\mu_2;\mu)}\!\otimes\!\kett{\mu}.
  \label{eq:QRW_boundary_state}
\end{align}
The ratio $S^{(k+1)}_{\rho\mu}/S^{(k+1)}_{0\mu}$ is the eigenvalue of the diagonal Verlinde line $X^{(k+1)}_\rho$ on sector $\mu$. The product of the two center generators multiplies each Ishibashi state by $(-1)^{\mu_1+\mu_2}=(-1)^\mu$. The identity $S^{(k+1)}_{k+1-\rho,\mu}=(-1)^\mu S^{(k+1)}_{\rho\mu}$ thus gives the center action $\rho\mapsto k+1-\rho$ used in Eq.~\eqref{eq:rho_star}.

Taking the overlap of two QRW states gives the closed-channel annulus
\begin{equation}
  Z_{BB'}(\tilde q)=\!\!\sum_{\substack{(\mu_1,\mu_2;\mu)\\ \mu_1+\mu_2+\mu\in2\mathbb Z}}\!\!
  \frac{S^{(k)}_{\rho_1\mu_1}S^{(k)}_{\rho_1'\mu_1}}{S^{(k)}_{0\mu_1}}\,
  \frac{S^{(1)}_{\rho_2\mu_2}S^{(1)}_{\rho_2'\mu_2}}{S^{(1)}_{0\mu_2}}\,
  \frac{S^{(k+1)}_{\rho\mu}S^{(k+1)}_{\rho'\mu}}{\big(S^{(k+1)}_{0\mu}\big)^2}\;
  \chi^{\rm cos}_{(\mu_1,\mu_2;\mu)}(\tilde q)\,\chi^{(k+1)}_{\mu}(\tilde q).
\end{equation}
Let $N^{(K)\,z}_{xy}$ denote the multiplicity of representation $z$ in the fusion of $x$ and $y$ at level $K$. We suppress the level when all labels belong to the same algebra. Applying~\eqref{eq:cosetS} and $\chi^{(k+1)}_{\mu}(\tilde q)=\sum_{\nu'}S^{(k+1)}_{\mu\nu'}\chi^{(k+1)}_{\nu'}(q)$, and implementing the selection rule on $(\mu_1,\mu_2;\mu)$ by the projector $\tfrac12[1+(-1)^{\mu_1+\mu_2+\mu}]$ together with the simple-current property of $S$, the $\mu_1$ and $\mu_2$ sums become Verlinde sums~\cite{Verlinde1988}, $\sum_{\mu_1}S_{\rho_1\mu_1}S_{\rho_1'\mu_1}S_{\mu_1\nu_1}/S_{0\mu_1}=N^{\nu_1}_{\rho_1\rho_1'}$, while the $\mu$ sum carries two factors of $1/S_{0\mu}$:
\begin{equation}
  \sum_{\mu}\frac{S_{\rho\mu}S_{\rho'\mu}}{S_{0\mu}}\,\frac{S_{\nu\mu}S_{\nu'\mu}}{S_{0\mu}}
  =\sum_{\mu}\Big(\sum_\lambda N^\lambda_{\rho\rho'}S_{\lambda\mu}\Big)\Big(\sum_{\lambda'}N^{\lambda'}_{\nu\nu'}S_{\lambda'\mu}\Big)
  =\sum_{\lambda}N^{\lambda}_{\rho\rho'}N^{\lambda}_{\nu\nu'} .
\end{equation}
The two projector terms are exchanged by the field identification, leaving one sum over all parity-allowed triples $(\nu_1,\nu_2;\nu)$. Here $\nu_1=0,\ldots,k$, $\nu_2=0,1$, $\nu=0,\ldots,k+1$, and $\nu_1+\nu_2+\nu$ is even. Both members of each identification class occur in this sum. The diagonal representation label $\nu'=0,\ldots,k+1$ is summed independently. The multiplicity $\mathcal N^{\nu\nu'}_{\rho\rho'}$ counts the intermediate labels $\lambda=0,\ldots,k+1$ that occur in both $\rho\otimes\rho'$ and $\nu\otimes\nu'$. In terms of these multiplicities we obtain
\begin{align}
  Z_{BB'}(q)&=\!\!\sum_{(\nu_1,\nu_2;\nu),\,\nu'}\!\!
  N^{(k)\,\nu_1}_{\rho_1\rho_1'}\,N^{(1)\,\nu_2}_{\rho_2\rho_2'}\,
  \mathcal N^{\nu\nu'}_{\rho\rho'}\,
  \chi^{\rm cos}_{(\nu_1,\nu_2;\nu)}(q)\,\chi^{(k+1)}_{\nu'}(q),
  \nonumber\\
  \mathcal N^{\nu\nu'}_{\rho\rho'}&=\sum_{\lambda}N^{(k+1)\,\lambda}_{\rho\rho'}\,N^{(k+1)\,\lambda}_{\nu\nu'} .
  \label{eq:annulus}
\end{align}
The two factors of $1/S_{0\mu}$ in the closed-channel annulus give the sum over $\lambda$, so the coset branching label $\nu$ and the diagonal label $\nu'$ can differ for a transmitting interface. For $\rho=\rho'=\rho_\ast$, the fusion $\rho_\ast\otimes\rho_\ast$ contains each even label once. The parity constraints carried by the fusion coefficients imply $\nu'\equiv(\rho_1+\rho_2+\rho)+(\rho_1'+\rho_2'+\rho')\pmod 2$. For $B=B'$ only even $\nu'$, i.e.\ integer total spin, appear, as appropriate for a ring with an even number of sites.

\paragraph*{Structure of the open channel.} By Eq.~\eqref{eq:line_fusion} and $X_\rho X_{\rho'}=\sum_\lambda N^{\lambda}_{\rho\rho'}X_\lambda$, the open channel of $B_{0,0;\rho}$ is the vacuum sector of $\suk{k}\otimes\suk{1}$ in which the $\suk{k+1}$ factor has been fused with every $\lambda\in\rho\otimes\rho$,
\begin{equation}
  Z_{BB}=\sum_{\lambda\in\rho\otimes\rho}Z^{(\lambda)},\qquad
  Z^{(\lambda)}=\sum_{\nu,\nu'}N^{(k+1)\,\nu'}_{\lambda\nu}\,\chi^{\rm cos}_{(0,0;\nu)}\,\chi^{(k+1)}_{\nu'} .
  \label{eq:Zlambda}
\end{equation}
For the undressed representative $B_{0,0;\rho_\ast}$, $N^{(k+1)\,\lambda}_{\rho_\ast\rho_\ast}=1$ for every even $\lambda$, so that $\mathcal N^{0\nu'}_{\rho_\ast\rho_\ast}=1$ for all even $\nu'$ and
\begin{equation}
  Z_{BB}(q)=\chi^{\rm cos}_{\rm id}(q)\!\!\sum_{\nu'\ \rm even}\!\!\chi^{(k+1)}_{\nu'}(q)\;+\;(\text{terms with }h^{\rm cos}>0).
  \label{eq:ZBB_structure}
\end{equation}

For anomaly obstructions to symmetric boundaries, see Refs.~\cite{JensenShaverinYarom2018,HanTiwariHsiehRyu2017}. The condition used here is invariance of the boundary state under the center action. The relation between this condition and endpoints of symmetry lines is discussed in Ref.~\cite{ChoiRayhaunSanghaviShao2023}, and its relation to mixed anomalies in Refs.~\cite{NumasawaYamaguchi2018,LiHsiehYaoOshikawa2024}. Equation~\eqref{eq:rho_star} is the interface counterpart of the $\mathbb Z_N$ selection rules of Ref.~\cite{FukusumiNakashiba2026}.

\paragraph*{Center parity of the defect operators.} Since the diagonal center line $J=X^{(k)}_k\otimes X^{(1)}_1$ leaves $B_{0,0;\rho_\ast}$ invariant, it acts on the open-channel Hilbert space $\mathcal H_{BB}$ and grades the defect operators. The twisted trace $\mathrm{Tr}_{\mathcal H_{BB}}J\,q^{L_0-c/24}$ must, after a modular transformation, become a non-negative combination of the characters of the $J$-twisted sector of the folded theory, in which the right-movers of the sector $(\mu_1,\mu_2)$ are replaced by those of $(k-\mu_1,1-\mu_2)$; for $k=3$ the $\mathcal A$-representations supporting twisted Ishibashi states are $(0,0;2)\otimes2$ and $(2,0;2)\otimes2$ only. Scanning all parity assignments of the open sectors of $B_{0,0;2}$ (allowing different parities for the three copies of $(0,0;2)\otimes2$) we find exactly one graded character:
\begin{equation}
  \mathrm{Tr}_{\mathcal H_{BB}}\,J\,q^{L_0-c/24}=\sum_{\lambda\in\rho_\ast\otimes\rho_\ast}(-1)^{\lambda/2}\,Z^{(\lambda)}(q),
  \label{eq:centre_parity}
\end{equation}
where the trace is represented through the fusion-channel decomposition~\eqref{eq:Zlambda}. The modular constraints fix eigenspace multiplicities: in the multiplicity-three sector $(0,0;2)\otimes2$ they fix two even copies and one odd copy, but not which fusion channel carries the odd one, nor that the action is diagonal in the fusion basis; for the sectors of multiplicity one, which include the relevant singlet and the lowest triplet, the parity is unambiguous. The modular transform of Eq.~\eqref{eq:centre_parity} is $3\sqrt{\tfrac{5-\sqrt5}{10}}\,\chi_{(0,0;2)}\chi_2+3\sqrt{\tfrac{5+\sqrt5}{10}}\,\chi_{(2,0;2)}\chi_2\simeq1.577\,\chi_{(0,0;2)}\chi_2+2.552\,\chi_{(2,0;2)}\chi_2$, with positive coefficients, as required. In matrix form, on $\nu,\nu'\in\{0,2,4\}$, the multiplicities and their signed version are $\mathcal N=\big(\begin{smallmatrix}1&1&1\\1&3&1\\1&1&1\end{smallmatrix}\big)$ and $\mathcal N_J=\big(\begin{smallmatrix}1&-1&1\\-1&1&-1\\1&-1&1\end{smallmatrix}\big)$. For $k=3$ the center-even sectors are therefore $(0,0;\nu)\otimes\nu'$ with $(\nu,\nu')\in\{(0,0),(4,4),(0,4),(4,0)\}$ and two of the three copies of $(2,2)$, with lowest weights $0,4,1,3,1,1$, while the center-odd ones are $(0,2),(2,0),(2,4),(4,2)$ and the remaining copy of $(2,2)$, with lowest weights $\tfrac13,\tfrac23,\tfrac53,\tfrac{10}3,1$. In particular the spin-one primary at $h=1/3$ and the singlet $\phi_{1,3}\otimes\chi_0$ at $h=2/3$ are odd, and the spin-two primary at $h=1$ is even. For $k=3$ this is consistent with the $g$-theorem~\cite{AffleckLudwig1991,FriedanKonechny2004}: if a center-even relevant singlet generated a nontrivial flow ending at a center-symmetric $\mathcal A$-preserving conformal boundary, the endpoint would have smaller $g$, which the classification excludes, since every such state is a non-negative integer combination of the QRW states and has $g\ge2g_{0\otimes0}=g_{B_{0,0;2}}$ (App.~\ref{app:dressing}); the direct evidence for the absence of such a singlet is the spectrum and its grading. The exhaustive scan of parity assignments singles out the same rule for $k=5$ (for $k=7$ the scan is no longer feasible), and it is in fact the unique graded character for every odd $k$, as shown at the end of this paragraph. With $N=k+1$ and $\rho_\ast=N/2$ one has $S^{(N)}_{\lambda\rho_\ast}=\sqrt{2/(N+2)}\,\sin\frac{\pi(\lambda+1)}{2}$, so that $\sum_{\lambda\ {\rm even}}(-1)^{\lambda/2}S^{(N)}_{\lambda x}=\sqrt{(N+2)/2}\;\delta_{x\rho_\ast}$ by unitarity of $S^{(N)}$, and the Verlinde formula gives the signed multiplicities in closed form,
\begin{equation}
  \mathcal N_J^{\nu\nu'}\equiv\sum_{\lambda\ {\rm even}}(-1)^{\lambda/2}N^{(N)\,\lambda}_{\nu\nu'}
  =\frac{N+2}{2}\,S^{(N)}_{\nu\rho_\ast}S^{(N)}_{\nu'\rho_\ast}=(-1)^{(\nu+\nu')/2}
  \qquad(\nu,\nu'\ {\rm even}),
  \label{eq:NJ_general}
\end{equation}
which reproduces the matrix $\mathcal N_J$ above for $k=3$. The same orthogonality localizes the modular transform of $\sum_{\nu,\nu'}\mathcal N_J^{\nu\nu'}\chi^{\rm cos}_{(0,0;\nu)}\chi^{(N)}_{\nu'}$ on the twisted Ishibashi sectors, both $\suk{N}$ labels being fixed to $\rho_\ast$:
\begin{equation}
  \widetilde Z_J=\sum_{[\mu_1,\mu_2;\rho_\ast]}(N+2)\,S^{(k)}_{0\mu_1}S^{(1)}_{0\mu_2}\,\chi^{\rm cos}_{(\mu_1,\mu_2;\rho_\ast)}\chi^{(N)}_{\rho_\ast},
\end{equation}
summed once over each identification class with $\mu_1+\mu_2+\rho_\ast$ even, with coefficients that are manifestly positive; for $k=3$ they are the two numbers quoted above. This establishes the modular consistency of the candidate trace for all odd $k$, and the relevant singlet of $B_{0,0;\rho_\ast}$, the coset primary $(0,0;2)$ with the affine vacuum, of weight $h_{1,3}=(k+1)/(k+3)$, appears only in the channel $\lambda=2$ and is center-odd under it. In fact the graded character is unique for every odd $k$, by inverting the modular transformation. Let $t_{\nu\nu'}$ be the trace of the center on the multiplicity space of the open sector $(0,0;\nu)\otimes\nu'$, with $t_{\nu\nu'}=0$ for odd labels and $t_{00}=1$ because the vacuum is invariant. Since the center acts trivially on the gluing algebra, the twined character is $\sum_{\nu,\nu'}t_{\nu\nu'}\chi^{\rm cos}_{(0,0;\nu)}\chi^{(N)}_{\nu'}$, and after the modular transformation the coefficient of $\chi^{\rm cos}_{(\mu_1,\mu_2;x)}\chi^{(N)}_{y}$ is $2S^{(k)}_{0\mu_1}S^{(1)}_{0\mu_2}(S^{(N)}tS^{(N)})_{xy}$. In the center-twisted bulk sector the left and right coset representations are related by the field identification while the affine representations must agree, which forces $x=y=N/2$; as the prefactor never vanishes, $S^{(N)}tS^{(N)}=C\,e_{N/2}e_{N/2}^{T}$. Using $(S^{(N)})^2=1$ and $t_{00}=1$ gives $C=(S^{(N)}_{0,N/2})^{-2}=(N+2)/2$ and hence Eq.~\eqref{eq:NJ_general}, which we have also checked numerically for $k\le13$. This fixes the dimensions of the two parity eigenspaces, $n^\pm=(n\pm t)/2$, but not which vector of a multiplicity space is odd in the fusion-channel basis; the explicit action on the fusion-junction spaces remains to be supplied.

This suffices for the stability of the undressed interface at every odd $k$. A relevant boundary field has $h<1$ and is therefore an affine primary at grade zero, and an $\mathrm{SU}(2)$ singlet must lie in the diagonal affine vacuum, $\nu'=0$; the coset primaries available there are $(0,0;2j)$ with weights $h_{1,2j+1}=j[(k+2)j-1]/(k+3)$, $j=0,1,\dots,(k+1)/2$, of which only $j=1$ gives a nonidentity relevant field. Its multiplicity is $\mathcal N^{20}_{\rho_\ast\rho_\ast}=1$, so its parity is the trace $t_{20}=-1$. At weight one the affine vacuum module contains only the current triplet, and the level-one Virasoro descendant of the coset vacuum vanishes, so there is no marginal singlet either. For every odd $k\ge3$, the only nonidentity relevant $\mathrm{SU}(2)$ singlet of the undressed interface is therefore $\phi_{1,3}\otimes1$, it is odd under the center, and the undressed interface and its invertible dressing are stable against local perturbations preserving both symmetries.

\paragraph*{Completeness.} The states~\eqref{eq:QRW_boundary_state} exhaust the elementary boundary states with the gluing~\eqref{eq:gluing}. Write a boundary state in the normalization $\ket{B}=\sum_i\psi_i\,\kett{i}/\sqrt{S^{\mathcal A}_{0i}}$, where $i$ runs over the Ishibashi states (with multiplicity) and $S^{\mathcal A}_{0i}=\sqrt2\,S^{(k)}_{0\mu_1}(S^{(k+1)}_{0\mu})^2$ is the vacuum row of the $\mathcal A$ modular matrix, $S^{\rm cos}_{0,[\mu]}S^{(k+1)}_{0\mu}$ with $S^{(1)}_{0\mu_2}=1/\sqrt2$. Using the identification to set $\rho_2=0$, Eq.~\eqref{eq:QRW_boundary_state} gives for the $(k+1)(k+2)$ states $\alpha=(\rho_1,\rho)$
\begin{equation}
  \psi_{(\rho_1,\rho),(\mu_1,\mu)}=S^{(k)}_{\rho_1\mu_1}S^{(k+1)}_{\rho\mu},\qquad \psi=S^{(k)}\otimes S^{(k+1)},
  \label{eq:psi}
\end{equation}
a unitary $(k+1)(k+2)$-dimensional matrix~\cite{QuellaSchomerus2002,Quella:2006de}. Let $X$ be any consistent boundary state with the same gluing, with coefficients $\psi_{Xi}$. Open--closed duality gives the vacuum multiplicity in the annulus between $B_\alpha$ and $X$ as $n^0_{\alpha X}=\sum_i\psi^*_{\alpha i}\psi_{Xi}\in\mathbb Z_{\ge0}$, and unitarity of $\psi$ then gives
\begin{equation}
  \ket X=\sum_\alpha n^0_{\alpha X}\,\ket{B_\alpha},\qquad n^0_{XX}=\sum_\alpha\big(n^0_{\alpha X}\big)^2 .
\end{equation}
If $X$ is elementary, $n^0_{XX}=1$, so exactly one $n^0_{\alpha X}$ equals one and all others vanish: $X=B_\alpha$. Every consistent boundary state with the gluing~\eqref{eq:gluing} is therefore a non-negative integer combination of the states~\eqref{eq:QRW_boundary_state}. A vacuum in the mixed open channel is a boundary-changing operator of dimension zero, which between elementary unitary boundaries is an isomorphism, so the result identifies the boundary conditions themselves and not only their states.

\paragraph*{Why the argument does not extend to $k'>1$.} For an interface between $\suk{k}$ and $\suk{k'}$ the first step goes through unchanged: the global $\mathrm{SU}(2)$ enforces the gluing of the diagonal $\suk{k+k'}$, and conformal invariance then implies the gluing of the coset stress tensor. The second step fails. For $k'=1$ the commutant of the diagonal currents is the Virasoro algebra of a minimal model, so these two conditions already amount to a rational gluing with finitely many Ishibashi states, and the unitarity argument above applies. When both levels exceed one, the coset $\suk{k}\otimes\suk{k'}/\suk{k+k'}$ has $c\ge1$ and its chiral algebra is larger than Virasoro. Preservation of this full algebra does not follow from the symmetry, and with respect to the algebra that is actually enforced, Virasoro times $\suk{k+k'}$, the product theory is not rational: there are infinitely many Ishibashi states, the coset Cardy states are no longer guaranteed to exhaust the elementary boundary conditions, and the center, which for even levels has fixed points among the factorizing states, need not single out one boundary label. A classification for $k'>1$ would require either a proof that the RG interface preserves the full coset algebra or control over the boundary states that break it; we do not attempt it here.

\paragraph*{Cardy consistency.} As an independent check, we have verified numerically for $k=3$ and $k=5$ (20 and 42 states, equal to the numbers of Ishibashi states) that the states~\eqref{eq:QRW_boundary_state}, labeled by all triples~\eqref{eq:labels} modulo the identification, yield non-negative integer multiplicities in all annuli $Z_{BB'}$, and that the identity multiplicity is $n^0_{BB'}=\delta_{BB'}$ for the inequivalent elementary states. The $g$-factor~\eqref{eq:gfactor} is consistent with the open channel through $g_B^2=\sum_R n^R_{BB}S^{\mathcal A}_{R,0}$, which we have also verified; for $k=3$, $g_{B_{0,0;2}}=1.0254$ and $g_{B_{2,0;2}}=1.6591$.

\paragraph*{Special cases.} For $\rho'=0$, $\mathcal N^{\nu\nu'}_{\rho0}=N^{\nu'}_{\rho\nu}$; in particular $\mathcal N^{\nu\nu'}_{00}=\delta_{\nu\nu'}$ and the labels are tied, $Z_{(\rho_1\rho_20),(\rho_1'\rho_2'0)}=\sum N^{\nu_1}_{\rho_1\rho_1'}N^{\nu_2}_{\rho_2\rho_2'}\chi^{(k)}_{\nu_1}\chi^{(1)}_{\nu_2}$, as it must be for factorizing boundary conditions. For $\rho=\rho'=\rho_\ast$ one has $N^{\lambda}_{\rho_\ast\rho_\ast}=1$ for every even $\lambda\le k+1$, so $\mathcal N^{\nu\nu'}_{\rho_\ast\rho_\ast}$ is the number of even labels in $\nu\otimes\nu'$. For $k=3$ this gives the matrix $\mathcal N^{\nu\nu'}=\big(\begin{smallmatrix}1&1&1\\1&3&1\\1&1&1\end{smallmatrix}\big)$ on $\nu,\nu'\in\{0,2,4\}$ and Eq.~\eqref{eq:ZBB_k3}, which we have also confirmed by direct evaluation of the modular transformation.

\paragraph*{Dressed states.} For $B_{\rho_1,0;\rho_\ast}$ the coset sectors $(\nu_1,0;\nu)$ with $\nu_1\in\rho_1\otimes\rho_1$ appear. All of them except $(0,0;\nu)$ have $h^{\rm cos}>0$ for every $\nu$, so the lowest level in each $S^z$ sector is unchanged, while excited levels are added. Since $\rho_1\otimes\rho_1=(k-\rho_1)\otimes(k-\rho_1)$, the dressings $\rho_1$ and $k-\rho_1$ have identical $Z_{BB}$. For $k=3$ and $\rho_1=1,2$ the additional sectors are $(2,0;\nu)$, $\nu=0,2,4$, with $h^{\rm cos}=7/5,\ 1/15,\ 2/5$, each paired with $\suk{4}$ characters with the multiplicities $\mathcal N^{\nu\nu'}$; the lightest, $(2,0;2)\otimes\chi_0$ at $h=1/15$, is an $\mathrm{SU}(2)$ singlet.

This singlet is present for every non-invertibly dressed transmitting representative. For $0<\rho_1<k$ and $1\le\rho\le k$ one has $N^{(k)\,2}_{\rho_1\rho_1}=1$, and $\mathcal N^{20}_{\rho\rho}=\sum_\lambda N^{(k+1)\,\lambda}_{\rho\rho}N^{(k+1)\,\lambda}_{20}\ge N^{(k+1)\,2}_{\rho\rho}N^{(k+1)\,2}_{20}=1$, so that Eq.~\eqref{eq:annulus} contains the sector $(2,0;2)\otimes0$, i.e.\ the coset primary $\phi_{3,3}$ of weight $h_{3,3}=2/[(k+2)(k+3)]$ with the affine vacuum. Its lowest state is an $\mathrm{SU}(2)$ singlet at
\begin{equation}
  \frac{h_{3,3}}{h_1}=\frac{1}{k+2},
  \label{eq:dressed_singlet}
\end{equation}
which we have verified directly from Eq.~\eqref{eq:annulus} for all $0<\rho_1<k$, $1\le\rho\le k$ and $k\le9$. For the factorizing labels $\rho=0,k+1$ the coset and affine labels are tied and this sector is absent, so the argument does not constrain dressed factorizing boundaries. Together with Eq.~\eqref{eq:dressing_test} this is what the first excited level of the $S^z=0$ sector tests.

\section{Magnetization-resolved weights of affine modules}
\label{app:weights}
The smallest $L_0$ eigenvalue among states with $J^3_0=m$ in the $\suk{N}$ module of highest weight $\nu'$ follows from the decomposition of $\suk{N}$ into $U(1)_{2N}$ and $\mathbb Z_N$-parafermion sectors~\cite{FateevZamolodchikov1985,GepnerQiu1987}, a branching with selection rules and identifications rather than an unconstrained tensor product. The $U(1)$ part contributes $m^2/N$, and the parafermion part is the weight of the primary $\Phi^{\nu'}_{2m}$,
\begin{equation}
  \varepsilon_N(\nu',m)=\frac{m^2}{N}+h^{\rm pf}_{(\nu',2m)},\qquad
  h^{\rm pf}_{(\ell,n)}=\frac{\ell(\ell+2)}{4(N+2)}-\frac{n^2}{4N}\quad(|n|\le\ell),
\end{equation}
extended to all $n$ by $n\sim n+2N$ and $(\ell,n)\sim(N-\ell,n\pm N)$.

\paragraph*{Lowest level and the Casimir bound.} By Eq.~\eqref{eq:ZBB_structure} the lowest level in the sector $S^z=m$ is
\begin{equation}
  h_m=\min_{\nu'\ \rm even}\ \varepsilon_{k+1}(\nu',m),
  \label{eq:hm_def}
\end{equation}
since the terms with $h^{\rm cos}>0$ have the same $\suk{k+1}$ content shifted upwards. For $2m\le k+1$ the minimum is attained by the highest-weight state of the spin-$m$ module, Eq.~\eqref{eq:casimir}: a state with $J^3_0=m$ has zero-mode spin $j\ge m$, and the Sugawara form of $L_0$, whose oscillator part $\frac{2}{k+3}\sum_{n>0}J^a_{-n}J^a_n$ is positive, gives $h\ge j(j+1)/(k+3)$ for every state of the preserved diagonal algebra.

\paragraph*{Closed form.}
For the candidate $\rho_\ast=N/2$ the minimum in Eq.~\eqref{eq:hm_def} runs over all even $\nu'$. Within the charge class $n=2m\ ({\rm mod}\ 2N)$ the parafermion weight $h^{\rm pf}_{(\ell,n)}$ decreases with decreasing $\ell$ at fixed $n$ down to $\ell=|n|$ (after reduction by the identifications), so the minimum is attained by the field $\Phi^{2d}_{\pm2d}$, where $2d\in[0,N/2]$ is the representative of $\pm2m$ obtained after using both identifications, i.e.\ $d={\rm dist}(m,\rho_\ast\mathbb Z)=\min(\bar m,\rho_\ast-\bar m)$, $\bar m=m\bmod\rho_\ast$. Its weight is
\begin{equation}
  \delta_m=h^{\rm pf}_{(2d,2d)}=\frac{d(d+1)}{N+2}-\frac{d^2}{N}
  =\frac{2d(\rho_\ast-d)}{N(N+2)}
  =\frac{2\bar m(\rho_\ast-\bar m)}{N(N+2)} ,
  \label{eq:delta_general}
\end{equation}
where the last step uses $d(\rho_\ast-d)=\bar m(\rho_\ast-\bar m)$, which holds for both branches of the minimum. For $N=k+1$ this is Eq.~\eqref{eq:delta_closed}; we have verified it against the direct minimization for $N\le12$. Setting $m=1$ gives $h_1=2/(N+2)$, and
\begin{equation}
  \frac{h_m}{h_1}=\frac{m^2(N+2)+2\bar m(\rho_\ast-\bar m)}{2N} ,
\end{equation}
which is an integer for all $m$ [write $m=q\rho_\ast+\bar m$ and expand]. For $0\le m\le\rho_\ast$ one has $\bar m(\rho_\ast-\bar m)=m(\rho_\ast-m)$ and $h_m/h_1=m(m+1)/2$, the triangular numbers of Eq.~\eqref{eq:casimir}; at $m=\rho_\ast+1$ the ratio is $(\rho_\ast+1)(\rho_\ast+2)/2+1$.

For $k=5$ Eq.~\eqref{eq:delta_closed} reads
\begin{equation}
  h_m=\frac{m^2}{6}+\frac{1}{12}\,\big[1-\delta_{3|m}\big],\qquad \frac{h_m}{h_1}=0,1,3,6,11,17,24,\dots
  \label{eq:hm_k5}
\end{equation}

\paragraph*{$N=4$ and $N=6$.} For $N=4$ the parafermion weights are $h^{\rm pf}_{(0,0)}=0$, $h^{\rm pf}_{(2,0)}=1/3$, $h^{\rm pf}_{(2,\pm2)}=1/12$, $h^{\rm pf}_{(4,0)}\sim h^{\rm pf}_{(0,4)}=1$, $h^{\rm pf}_{(4,\pm2)}\sim h^{\rm pf}_{(0,\mp2)}=3/4$, $h^{\rm pf}_{(4,\pm4)}\sim h^{\rm pf}_{(0,0)}=0$. Table~\ref{tab:eps} lists $\varepsilon_4(\nu',m)$ and $\varepsilon_6(\nu',m)$ for the even modules. The minimum over even $\nu'$ gives Eqs.~\eqref{eq:hm_k3} and~\eqref{eq:hm_k5}; the minimum of the sum of the coset and affine weights over the allowed sectors for the undressed representatives $B_{0,0;\rho}$ gives the corresponding rows of Table~\ref{tab:results} (in the coset vacuum sector $\nu'\in\rho\otimes\rho$; for general $\nu$, $\nu'$ occurs whenever some $\lambda\in\rho\otimes\rho$ has $N^{(k+1)\,\nu'}_{\lambda\nu}>0$); for other values of the boundary label $\rho$ a non-invertible UV dressing can lower these minima, because the extra coset sectors it adds carry charged states of smaller total weight. The states realizing the minimum for $k=3$ are: $m=1$, $\Phi^{(2)}_{S^z=1}$ (spin-one primary, $h=1/3$); $m=2$, $\Phi^{(4)}_{S^z=2}$ (spin-two primary, $h=1$); $m=3$, $J^+_{-1}J^+_{-1}\Phi^{(2)}_{S^z=1}$ ($h=7/3$); $m=4$, $(J^+_{-1})^4|0\rangle$ ($h=4$); $m=5$, $(J^+_{-2})^2(J^+_{-1})^2\Phi^{(2)}_{S^z=1}$ ($h=19/3$; the state $(J^+_{-1})^3\Phi^{(2)}$ is null in the integrable spin-one module of $\suk{4}$).

For the undressed spectrum, the sectors with $h^{\rm cos}>0$ in~\eqref{eq:ZBB_structure} cannot produce the lowest level: for the same $\suk{k+1}$ content they are shifted upward by $h^{\rm cos}\ge h_{(1,3)}$, and the $\suk{k+1}$ content of all sectors in~\eqref{eq:ZBB_k3} is the same set of even modules.

\begin{table}[h]
\centering
\begin{ruledtabular}
\begin{tabular}{c|cccccc||c|ccccccc}
$\suk{4}$ & $m\!=\!0$ & 1 & 2 & 3 & 4 & 5 & $\suk{6}$ & $m\!=\!0$ & 1 & 2 & 3 & 4 & 5 & 6\\
\hline
$\nu'=0$ & $\mathbf{0}$ & 1 & 2 & 3 & $\mathbf{4}$ & 7 & $\nu'=0$ & $\mathbf 0$ & 1 & 2 & 3 & 4 & 5 & $\mathbf 6$\\
$\nu'=2$ & $1/3$ & $\mathbf{1/3}$ & $4/3$ & $\mathbf{7/3}$ & $13/3$ & $\mathbf{19/3}$ & $\nu'=2$ & $1/4$ & $\mathbf{1/4}$ & $5/4$ & $9/4$ & $13/4$ & $\mathbf{17/4}$ & $25/4$\\
$\nu'=4$ & 1 & 1 & $\mathbf{1}$ & 3 & 5 & 7 & $\nu'=4$ & $3/4$ & $3/4$ & $\mathbf{3/4}$ & $7/4$ & $\mathbf{11/4}$ & $19/4$ & $27/4$\\
 & & & & & & & $\nu'=6$ & $3/2$ & $3/2$ & $3/2$ & $\mathbf{3/2}$ & $7/2$ & $11/2$ & $15/2$\\
\end{tabular}
\end{ruledtabular}
\caption{Lowest weight $\varepsilon_N(\nu',m)$ with $J^3_0=m$ in the even modules $\nu'$ of $\suk{4}$ (left) and $\suk{6}$ (right). Bold entries are the minima over even $\nu'$, Eqs.~\eqref{eq:hm_k3} and~\eqref{eq:hm_k5}.}
\label{tab:eps}
\end{table}

\section{Magnetization-resolved defect partition functions}
\label{app:lowlying}
The predictions of the main text can be read off directly from the open-channel partition function refined by the magnetization,
\begin{equation}
  Z_{BB}(q,y)=\mathrm{Tr}_{\mathcal H_{BB}}\,q^{L_0-c_A/24}\,y^{S^z},\qquad c_A=c_{\rm UV}+c_{\rm IR},
  \label{eq:Zqy}
\end{equation}
obtained from Eq.~\eqref{eq:annulus} with the Rocha-Caridi characters of $\mathcal M(k+2,k+3)$ and the $S^z$-graded Weyl--Kac characters of $\suk{k+1}$~\cite{DiFrancesco1997}. After removing the vacuum prefactor, $\widehat Z_{BB}(q,y)=q^{c_A/24}Z_{BB}(q,y)$, the smallest exponent multiplying $y^{m}$ in $\widehat Z_{BB}$ is $h_m$, and the normalized gap of Eq.~\eqref{eq:triangular} is $h_m/h_1$. The first terms are listed below; for the undressed state and its invertible center image the exponents are integer multiples of $h_1=2/(k+3)$, Eq.~\eqref{eq:tower}, and the coefficient of $q^{h}$ is the $\mathrm{SU}(2)$ character of the states at that level. The commensurability holds for every odd $k$: the open sectors are $(0,0;2j)\otimes2\ell$ with coset weight $h_{1,2j+1}=j[j(k+2)-1]/(k+3)$ and affine weight $\ell(\ell+1)/(k+3)$, so that for the primaries
\begin{equation}
  \frac{h}{h_1}=\frac{j\,[j(k+2)-1]+\ell(\ell+1)}{2}\in\mathbb Z ,
\end{equation}
both terms in the numerator being even, and descendants add integers, which are multiples of $h_1$ because $1/h_1=(k+3)/2$ is an integer. Non-invertible dressings introduce additional coset sectors $(\nu_1,0;\nu)$ with $\nu_1\neq0$, whose spectra need not remain commensurate with $h_1$: every dressing with $0<\rho_1<k$ contains $\nu_1=2$ and hence the singlet $(2,0;2)\otimes0$, i.e.\ $\phi_{3,3}\otimes\chi_0$ of weight $h_{3,3}=2/[(k+2)(k+3)]$, at $h/h_1=1/(k+2)$, so that its full spectrum is not contained in $h_1\mathbb Z$ [Eq.~\eqref{eq:ZB} for $k=3$]. Some of the additional sectors may nevertheless lie on the lattice; for $k=13$ and $\rho_1=5$, for instance, $(10,0;10)\otimes0$ has $h=h_{11,11}=h_1$.

\paragraph*{$k=3$, undressed interface $B_{0,0;2}$ ($c_A=14/5$, $h_1=1/3$).}
\begin{align}
  Z_{B_{0,0;2}}(q,y)&=q^{-7/60}\Big[\,1\notag\\
  &+(y^{-1}+1+y)\,q^{1/3}\notag\\
  &+q^{2/3}\notag\\
  &+(y^{-2}+5y^{-1}+5+5y+y^{2})\,q\notag\\
  &+(y^{-2}+2y^{-1}+3+2y+y^{2})\,q^{4/3}\notag\\
  &+(y^{-2}+2y^{-1}+3+2y+y^{2})\,q^{5/3}\notag\\
  &+(5y^{-2}+13y^{-1}+18+13y+5y^{2})\,q^{2}\notag\\
  &+(y^{-3}+3y^{-2}+7y^{-1}+8+7y+3y^{2}+y^{3})\,q^{7/3}\notag\\
  &\;+\cdots\Big]. \label{eq:ZA}
\end{align}
The lowest powers of $q$ multiplying $y,y^2,y^3$ are $q^{1/3},q^{1},q^{7/3}$, giving $h_m/h_1=1,3,7$; the $y^{4}$ and $y^{5}$ terms first appear at $q^{4}$ and $q^{19/3}$ (not displayed), giving $12$ and $19$. The decomposition into $\mathrm{SU}(2)$ multiplets is listed in Table~\ref{tab:lowlying}.

\paragraph*{$k=3$, Fibonacci dressings $B_{2,0;2}$ and $B_{1,0;2}$ (equal $Z_{BB}$).}
\begin{align}
  Z_{B_{2,0;2}}(q,y)&=q^{-7/60}\Big[\,1\notag\\
  &+q^{1/15}\notag\\
  &+(y^{-1}+1+y)\,q^{1/3}\notag\\
  &+(3y^{-1}+4+3y)\,q^{2/5}\notag\\
  &+q^{2/3}\notag\\
  &+(y^{-1}+1+y)\,q^{11/15}\notag\\
  &+(y^{-2}+5y^{-1}+5+5y+y^{2})\,q\notag\\
  &+(y^{-2}+2y^{-1}+3+2y+y^{2})\,q^{16/15}\notag\\
  &+(y^{-2}+2y^{-1}+3+2y+y^{2})\,q^{4/3}\notag\\
  &+(4y^{-2}+11y^{-1}+16+11y+4y^{2})\,q^{7/5}\notag\\
  &+(y^{-2}+2y^{-1}+3+2y+y^{2})\,q^{5/3}\notag\\
  &+(y^{-2}+4y^{-1}+5+4y+y^{2})\,q^{26/15}\notag\\
  &\;+\cdots\Big]. \label{eq:ZB}
\end{align}
The terms with exponents in $\tfrac13\mathbb Z$ coincide with those of Eq.~\eqref{eq:ZA}; that the lowest level in every $S^z$ sector, and hence the ratios $0,1,3,7,12,19$, are the same for the two states follows from the coset-positivity argument of App.~\ref{app:annulus}, since the additional sectors have $h^{\rm cos}>0$. The additional terms, whose exponents lie in $\tfrac1{15}+\tfrac13\mathbb Z_{\ge0}$ (note $\tfrac25=\tfrac1{15}+\tfrac13$), come from the coset sectors $(2,0;\nu)$ with $h^{\rm cos}=1/15,\,7/5,\,2/5$ (Table~\ref{tab:lowlying}). The lightest of them is the $S^z=0$ singlet at $q^{1/15}$, one fifth of the gap $h_1$; it is the only level below the first triplet, so the second level of the $S^z=0$ sector distinguishes the two states; the lattice value, Eq.~\eqref{eq:dressing_test}, disfavors the dressed states.

\paragraph*{$k=5$, $B_{0,0;3}$ ($c_A=22/7$, $h_1=1/4$).}
\begin{align}
  Z_{B_{0,0;3}}(q,y)&=q^{-11/84}\Big[\,1\notag\\
  &+(y^{-1}+1+y)\,q^{1/4}\notag\\
  &+(y^{-2}+y^{-1}+2+y+y^{2})\,q^{3/4}\notag\\
  &+(4y^{-1}+4+4y)\,q\notag\\
  &+(y^{-2}+2y^{-1}+3+2y+y^{2})\,q^{5/4}\notag\\
  &+(y^{-3}+4y^{-2}+4y^{-1}+4+4y+4y^{2}+y^{3})\,q^{3/2}\notag\\
  &+(y^{-3}+2y^{-2}+4y^{-1}+5+4y+2y^{2}+y^{3})\,q^{7/4}\notag\\
  &+(4y^{-2}+11y^{-1}+16+11y+4y^{2})\,q^{2}\notag\\
  &+(2y^{-3}+4y^{-2}+8y^{-1}+9+8y+4y^{2}+2y^{3})\,q^{9/4}\notag\\
  &+(4y^{-3}+11y^{-2}+14y^{-1}+14+14y+11y^{2}+4y^{3})\,q^{5/2}\notag\\
  &+(y^{-4}+3y^{-3}+8y^{-2}+12y^{-1}+16+12y+8y^{2}+3y^{3}+y^{4})\,q^{11/4}\notag\\
  &\;+\cdots\Big]. \label{eq:Zk5}
\end{align}
The lowest powers multiplying $y,y^2,y^3,y^4$ are $q^{1/4},q^{3/4},q^{3/2},q^{11/4}$, i.e.\ $h_m/h_1=1,3,6,11$: the spin-one, spin-two and spin-three primaries of $\suk{6}$ (triangular numbers), followed by the first non-integrable sector, $m=4$, where the ratio exceeds the triangular number $10$ by one. Note that the multiple $2h_1=1/2$ is not occupied, so the spectrum is commensurate with $h_1$ without being an equally spaced tower.

\paragraph*{$k=7$, $B_{0,0;4}$ ($c_A=10/3$, $h_1=1/5$).}
\begin{align}
  Z_{B_{0,0;4}}(q,y)&=q^{-5/36}\Big[\,1\notag\\
  &+(y^{-1}+1+y)\,q^{1/5}\notag\\
  &+(y^{-2}+y^{-1}+1+y+y^{2})\,q^{3/5}\notag\\
  &+q^{4/5}\notag\\
  &+(4y^{-1}+4+4y)\,q\notag\\
  &+(y^{-3}+2y^{-2}+3y^{-1}+4+3y+2y^{2}+y^{3})\,q^{6/5}\notag\\
  &+(3y^{-2}+3y^{-1}+3+3y+3y^{2})\,q^{7/5}\notag\\
  &+(y^{-3}+2y^{-2}+3y^{-1}+3+3y+2y^{2}+y^{3})\,q^{8/5}\notag\\
  &+(y^{-1}+2+y)\,q^{9/5}\notag\\
  &+(y^{-4}+4y^{-3}+8y^{-2}+15y^{-1}+20+15y+8y^{2}+4y^{3}+y^{4})\,q^{2}\notag\\
  &+(y^{-4}+3y^{-3}+6y^{-2}+10y^{-1}+11+10y+6y^{2}+3y^{3}+y^{4})\,q^{11/5}\notag\\
  &+(3y^{-3}+9y^{-2}+12y^{-1}+12+12y+9y^{2}+3y^{3})\,q^{12/5}\notag\\
  &+(y^{-4}+3y^{-3}+7y^{-2}+9y^{-1}+10+9y+7y^{2}+3y^{3}+y^{4})\,q^{13/5}\notag\\
  &+(y^{-4}+y^{-3}+2y^{-2}+4y^{-1}+7+4y+2y^{2}+y^{3}+y^{4})\,q^{14/5}\notag\\
  &+(4y^{-4}+15y^{-3}+28y^{-2}+50y^{-1}+58+50y+28y^{2}+15y^{3}+4y^{4})\,q^{3}\notag\\
  &+(y^{-5}+4y^{-4}+10y^{-3}+18y^{-2}+26y^{-1}+30+26y+18y^{2}+10y^{3}+4y^{4}+y^{5})\,q^{16/5}\notag\\
  &\;+\cdots\Big]. \label{eq:Zk7}
\end{align}
Here the lowest powers multiplying $y,\dots,y^5$ are $q^{1/5},q^{3/5},q^{6/5},q^{2},q^{16/5}$, i.e.\ $h_m/h_1=1,3,6,10,16$: the spin-one to spin-four primaries of $\suk{8}$ and then the first non-integrable sector $m=5$, where $16=15+1$.

\paragraph*{Multiplet content for $k=3$.} In terms of $\mathrm{SU}(2)$ multiplets $[S]$ the first levels of $B_{0,0;2}$ are
\begin{equation}
  \frac{h}{h_1}=0\,[0],\ \ 1\,[1],\ \ 2\,[0],\ \ 3\,\big(4[1]+[2]\big),\ \ 4\,\big([0]+[1]+[2]\big),\ \dots
  \label{eq:excited_k3}
\end{equation}
The first singlet excitation, at $h/h_1=2$, is the coset primary $\phi_{1,3}$ of $\mathcal M(5,6)$ with $\chi_0$ of $\suk{4}$, the $\mathrm{SU}(2)$-neutral excitation carried by the interface; it is odd under the center, Eq.~\eqref{eq:centre_parity}.
Table~\ref{tab:lowlying} lists the open-channel spectrum of $B_{0,0;2}$ up to $h=7h_1$, decomposed into $\mathrm{SU}(2)$ multiplets, together with the additional levels of the dressed states, none of which is a multiple of $h_1$.

\begin{table}[h]
\centering
\begin{ruledtabular}
\begin{tabular}{cl|cl}
\multicolumn{2}{c|}{$B_{0,0;2}$} & \multicolumn{2}{c}{additional levels of $B_{1,0;2}$, $B_{2,0;2}$}\\
$h/h_1$ & content & $h/h_1$ & content \\
\hline
0 & $[0]$ & $1/5$ & $[0]$ \\
1 & $[1]$ & $6/5$ & $[0]+3[1]$ \\
2 & $[0]$ & $11/5$ & $[1]$ \\
3 & $4[1]+[2]$ & $16/5$ & $[0]+[1]+[2]$ \\
4 & $[0]+[1]+[2]$ & $21/5$ & $5[0]+7[1]+4[2]$ \\
5 & $[0]+[1]+[2]$ & $26/5$ & $[0]+3[1]+[2]$ \\
6 & $5[0]+8[1]+5[2]$ & $31/5$ & $3[0]+3[1]+3[2]$ \\
7 & $[0]+4[1]+2[2]+[3]$ & & \\
\end{tabular}
\end{ruledtabular}
\caption{Low-lying open-channel spectrum for $k=3$ in units of $h_1=1/3$, with $\mathrm{SU}(2)$ content; $n[S]$ denotes $n$ multiplets of spin $S$. Left: undressed interface $B_{0,0;2}$ (and $B_{3,0;2}$). Right: additional levels of the dressed states $B_{1,0;2}$ and $B_{2,0;2}$.}
\label{tab:lowlying}
\end{table}

The lowest levels in each $S^z$ sector, which determine Eq.~\eqref{eq:hm_k3}, are the highest-weight states of $[1]$ at $h/h_1=1$, of $[2]$ at $3$, of $[3]$ at $7$, and so on. The four lowest distinct energy levels in the sectors $S^z=0,1,2,3$ are, in units of $h_1$, $\{0,1,2,3\}$, $\{1,3,4,5\}$, $\{3,4,5,6\}$, and $\{7,9,10,11\}$ for $B_{0,0;2}$, and $\{0,\tfrac15,1,\tfrac65\}$, $\{1,\tfrac65,\tfrac{11}5,3\}$, $\{3,\tfrac{16}5,4,\tfrac{21}5\}$, and $\{7,\tfrac{36}5,\tfrac{41}5,9\}$ for the dressed states.

\section{Transmission coefficients of the candidate interfaces}
\label{app:transmission}
QRW~\cite{Quella:2006de} express the reflection and transmission coefficients of a conformal interface through the normalized overlap of the boundary state with the spin-two quasi-primary $\Omega=c_{\rm IR}T_{\rm UV}-c_{\rm UV}T_{\rm IR}$ of the folded theory, where here $c_{\rm UV}=c_k=3k/(k+2)$ and $c_{\rm IR}=1$. With $\ket\Omega=(c_{\rm IR}L^{\rm UV}_{-2}-c_{\rm UV}L^{\rm IR}_{-2})\ket0$, $\langle\Omega|\Omega\rangle=\tfrac12c_{\rm UV}c_{\rm IR}(c_{\rm UV}+c_{\rm IR})$, and
\begin{equation}
  \omega_B=\frac{(\bra\Omega\otimes\bra{\bar\Omega})\ket B}{\tfrac12c_{\rm UV}c_{\rm IR}(c_{\rm UV}+c_{\rm IR})\,\langle0|B\rangle},
\end{equation}
they obtain $\mathcal T=\frac{2c_{\rm UV}c_{\rm IR}}{(c_{\rm UV}+c_{\rm IR})^2}(1-\omega_B)$; in the normalization of Refs.~\cite{Meineri2020,Furuta2025} the transmission coefficient is $c_{LR}=\frac{c_{\rm UV}c_{\rm IR}}{c_{\rm UV}+c_{\rm IR}}(1-\omega_B)$. For the states~\eqref{eq:QRW_boundary_state} we evaluate $\omega_B$ by decomposing $\Omega$ into representations of $\mathcal A$. At level two in the vacuum sector of $\suk{k}\otimes\suk{1}$ there are three $\mathrm{SU}(2)$ singlets, $L^{(k)}_{-2}$, $L^{(1)}_{-2}$, and $J_{(k),-1}\!\cdot\!J_{(1),-1}$ acting on the vacuum. Two of them, the Sugawara and coset stress tensors $T_{k+1}$ and $T_{\rm cos}$, belong to the identity representation $(0,0;0)\otimes0$ of $\mathcal A$. The third is the singlet descendant $J_{{\rm diag},-1}\!\cdot\!\Phi_{-1}$ of the $\suk{k+1}$ spin-one primary $\Phi^a=J^a_{(k)}-kJ^a_{(1)}$, which belongs to $(0,0;2)\otimes2$. Writing $\Omega=\alpha\,\Omega_{\mathcal A}+\beta\,J_{\rm diag}\!\cdot\!\Phi$ with $\Omega_{\mathcal A}=c_{\rm cos}T_{k+1}-c_{k+1}T_{\rm cos}$, the fraction of the norm of $\Omega$ in the identity representation is
\begin{equation}
  p=\frac{\alpha^2\|\Omega_{\mathcal A}\|^2}{\|\Omega\|^2}
  =\frac{c_k\,(k-1)^2}{(k+3)^2\,c_{\rm cos}\,c_{k+1}}
  =\frac{(k-1)^2}{(k+1)(k+5)},
  \label{eq:p}
\end{equation}
which we have also confirmed from the explicit Gram matrix of the three level-two states. The one-point coefficients of the two $\mathcal A$-Ishibashi states in~\eqref{eq:QRW_boundary_state} give
\begin{equation}
  \omega_B=p+(1-p)\,r_\rho,\qquad
  r_\rho=\frac{S^{(k+1)}_{\rho2}\,S^{(k+1)}_{00}}{S^{(k+1)}_{02}\,S^{(k+1)}_{\rho0}}
  =\frac{4\cos^2\!\frac{\pi(\rho+1)}{k+3}-1}{4\cos^2\!\frac{\pi}{k+3}-1}.
  \label{eq:omega_general}
\end{equation}
Using $\frac{c_k}{c_k+1}(1-p)=\frac{6k}{(k+1)(k+5)}$, the transmission coefficient of the candidate $\rho$ is
\begin{equation}
  c_{LR}(\rho)=\frac{6k}{(k+1)(k+5)}\,\big(1-r_\rho\big).
  \label{eq:cLR_general}
\end{equation}
For $\rho=0$ and $\rho=k+1$, $r_\rho=1$ and $\omega_B=1$: the interface factorizes. For $\rho_\ast=(k+1)/2$, $\cos\frac{\pi(\rho_\ast+1)}{k+3}=0$ and $r_{\rho_\ast}=-\big(4\cos^2\frac{\pi}{k+3}-1\big)^{-1}$, which is the minimum of $r_\rho$ and hence the maximum of the transmission; this gives Eq.~\eqref{eq:cLR_main}. Explicitly,
\begin{align}
  k=3:&\quad p=\tfrac18,\quad r_2=-\tfrac12,\quad \omega_B=-\tfrac{5}{16},\quad
  \mathcal T=\tfrac{135}{224},\quad c_{LR}=\tfrac{27}{32}\simeq0.844,\\
  k=5:&\quad p=\tfrac{4}{15},\quad r_3=1-\sqrt2,\quad \omega_B=1-\tfrac{11\sqrt2}{15},\quad
  \mathcal T=\tfrac{7\sqrt2}{22},\quad c_{LR}=\tfrac{1}{\sqrt2}\simeq0.707.
\end{align}
For $k=7,9,11$ one finds $c_{LR}=7(5-\sqrt5)/32\simeq0.6046$, $27(1+\sqrt3)/140\simeq0.5269$, and $0.4664$, so that for this family $c_{LR}$ decreases with $k$ while $c_{\rm UV}-c_{\rm IR}$ increases. For comparison, the candidates $\rho=1,3$ at $k=3$ have $c_{LR}=9/16$. All values respect $c_{LR}\le\min(c_{\rm UV},c_{\rm IR})=1$~\cite{Karch2024}.

\section{Topological dressing, interface entropy, and phantom currents}
\label{app:dressing}
\paragraph*{Dressings and $g$-factor.} For $\rho=\rho_\ast$ the labels $(\rho_1,\rho_2)$ remain free. Using the identification we can set $\rho_2=0$, which leaves $k+1$ inequivalent choices $\rho_1=0,\dots,k$. They correspond to fusing the Verlinde lines of the UV theory onto the interface, $B_{\rho_1,0;\rho_\ast}=X^{(k)}_{\rho_1}\star B_{0,0;\rho_\ast}$, i.e.\ to topological dressings. The dressing multiplies the $g$-factor~\cite{AffleckLudwig1991} by the quantum dimension of $\rho_1$:
\begin{equation}
  g_{B_{\rho_1,0;\rho_\ast}}
  =\langle0|B_{\rho_1,0;\rho_\ast}\rangle
  =\frac{S^{(k)}_{\rho_10}}{S^{(k)}_{00}}\;\frac{1}{\sin\frac{\pi}{k+3}}\;
  \sqrt{S^{(k)}_{00}S^{(1)}_{00}} ,
  \label{eq:gfactor}
\end{equation}
where $\sqrt{S^{(k)}_{00}S^{(1)}_{00}}$ is the $g$-factor of the factorizing state $\ket{0}\otimes\ket{0}$ and $S^{(k+1)}_{\rho_\ast0}/S^{(k+1)}_{00}=1/\sin\frac{\pi}{k+3}$ ($=2$ for $k=3$). The two dressings with the smallest $g$ are $\rho_1=0$ and $\rho_1=k$. They are exchanged by the center line acting on one side of the interface, $B_{k,0;\rho_\ast}=X^{(k)}_k\star B_{0,0;\rho_\ast}=B_{0,0;\rho_\ast}\star X^{(1)}_1$; a relative one-site shift of the microscopic junction is a possible realization of this change of representative, although establishing the correspondence would require an explicit comparison of the junction Hamiltonians. They have the same $g$-factor, the same transmission coefficient, and the same spectrum $Z_{BB}$, and we do not distinguish them. The remaining dressings $0<\rho_1<k$ have larger $g$; for $k=3$, $\rho_1=1,2$ give $g/g_{B_{0,0;2}}=(1+\sqrt5)/2$. The classification of the main text leaves $\rho_1$ undetermined, so the dressing has to be fixed by an additional dynamical criterion, such as the excited-state spectrum; the $k=3$ data provide a finite-size test of this choice. For $k=3$ every center-symmetric non-negative integer combination of QRW states has $g\ge2g_{0\otimes0}=g_{B_{0,0;2}}$: the symmetric elementary states are $B_{\rho_1,0;2}$ with $g=2d_{\rho_1}g_{0\otimes0}$, and a state with $\rho\neq2$ can only be symmetric as a sum with its center image, which doubles $g$, the smallest such pair being $\ket0\otimes\ket0\oplus\ket3\otimes\ket1$ with $g=2g_{0\otimes0}$. By the $g$-theorem~\cite{AffleckLudwig1991,FriedanKonechny2004,Casini2016g}, $B_{0,0;2}$ therefore admits no nontrivial symmetry-preserving boundary flow to another conformal boundary condition in this class, consistent with the absence of center-even relevant singlets found in App.~\ref{app:annulus}. This bound does not extend to $k>3$, where $g_{B_{0,0;\rho_\ast}}=g_{0\otimes0}/\sin\frac{\pi}{k+3}$ exceeds the $g=2g_{0\otimes0}$ of the symmetric factorizing pair ($2.613\,g_{0\otimes0}$ versus $2g_{0\otimes0}$ for $k=5$); there the stability follows from the spectrum and the uniqueness of the graded character (App.~\ref{app:annulus}), while the boundary-entropy comparison above is specific to $k=3$. That an RG interface should be the elementary interface with the smallest $g$-factor compatible with the symmetries is a natural expectation, though not a theorem~\cite{Gaiotto:2012np,KonechnySchmidtColinet2014}; the first excited state of the $S^z=0$ sector tests it directly, Eq.~\eqref{eq:dressing_test}.

\paragraph*{Relation to the phantom-current construction.} A general construction of RG interfaces between rational CFTs has recently been formulated in terms of the topological lines transmitted through the interface~\cite{Chen2026a}, building on Gaiotto's construction~\cite{Gaiotto:2012np} and on the phantom currents of Refs.~\cite{Antinucci2025,Furuta2025,Zhang2026}. A transmitted line carries chiral endpoint operators on both sides; after folding, their product is a chiral field of the product theory $\mathcal T_A$, a phantom current, and condensing the phantom currents of integer spin extends $\mathcal T_A$ to an auxiliary rational theory $\mathcal T_B$. In this construction, candidate RG interfaces are obtained by pulling automorphism-twisted Cardy states of $\mathcal T_B$ back to $\mathcal T_A$ through the topological interface of the extension. The interfaces classified here are consistent with this scheme, and the comparison is instructive; the classification itself does not rely on it.

For the flow~\eqref{eq:flow} the only nontrivial Verlinde line of $\suk{k}$ that commutes with $\Phi^{(1)}$ is the center line $X^{(k)}_k$ (the condition $S_{\ell2}/S_{02}=S_{\ell0}/S_{00}$ fails for $0<\ell<k$), matched in the IR to $X^{(1)}_1$. Its endpoint fields have weights $k/4$ and $1/4$, so the folded phantom current has spin $(k+1)/4$: integer for $k\equiv3\pmod4$, half-integer for $k\equiv1\pmod4$. For $k=3$ it is a spin-one current in the sector $V^{(3)}_3\otimes V^{(1)}_1$, with eight components (spin $\tfrac32\otimes\tfrac12=2\oplus1$). Adding them to the six currents of $\suk{3}\oplus\suk{1}$ gives fourteen: the extension is the conformal embedding
\begin{equation}
  \suk{1}\oplus\suk{3}\subset(G_2)_1,\qquad c=\tfrac{14}{5},
  \label{eq:G2}
\end{equation}
of Refs.~\cite{SchellekensWarner1986,BaisBouwknegt1987}, under which the $(G_2)_1$ vacuum branches as $V^{(3)}_0\otimes V^{(1)}_0\oplus V^{(3)}_3\otimes V^{(1)}_1$ and its only other primary, the seven-dimensional representation with $h=2/5$, as $V^{(3)}_2\otimes V^{(1)}_0\oplus V^{(3)}_1\otimes V^{(1)}_1$. The preserved diagonal $\mathrm{SU}(2)$ requires the gluing automorphism of $(G_2)_1$ to fix the diagonal $\suk{4}$ pointwise. This $\mathfrak{su}(2)$ has embedding index four in $G_2$ and is conjugate to $\mathfrak{so}(3)\subset\mathfrak{su}(3)\subset G_2$; since the adjoint of $G_2$ decomposes under it as $3\oplus3\oplus3\oplus5$, without singlets, its centralizer is finite. The centralizer contains the center $\mathbb Z_3$ of $\mathrm{SU}(3)$ and the center line, which together generate an $S_3$ acting on the two spin-one currents orthogonal to $J_{\rm diag}$, one of which, $\Phi^a=J^a_{(3)}-3J^a_{(1)}$, lies in $\suk{3}\oplus\suk{1}$ and the other in the phantom sector. The $\mathbb Z_3$ rotations mix $\Phi^a$ with the spin-one
phantom currents. Through the Sugawara construction, they
therefore rotate $T_{\rm UV}$ and $T_{\rm IR}$ into
conformal-weight-two $\mathrm{SU}(2)$-singlet combinations
of currents, while preserving $\mathcal A$ pointwise. The corresponding twisted Cardy states of $(G_2)_1$, pulled back to $\mathcal T_A$, provide candidates for the transmitting QRW interfaces (the classification guarantees only that a pull-back is a non-negative integer combination of QRW states); the reflection along the phantom direction is the quantum $\mathbb Z_2$ of the extension, which fixes $J_{(1)}$ and $J_{(3)}$ separately and therefore transmits no energy.

Two checks support the identification of these pull-backs with our interfaces. First, their $g$-factors are $\sqrt{S^B_{00}/S^A_{00}}\,S^B_{b0}/\sqrt{S^B_{00}}=2d_b\,g_{0\otimes0}$ for the two $(G_2)_1$ labels $b$, with $d_b=1$ and $(1+\sqrt5)/2$, which are exactly the $g$-factors~\eqref{eq:gfactor} of $B_{0,0;2}$ and of the dressed states $B_{1,0;2}$, $B_{2,0;2}$. Second, a pull-back from $(G_2)_1$ can have Ishibashi components only in center-even sectors $\mu_1+\mu_2\in2\mathbb Z$; indeed the coefficients~\eqref{eq:QRW_boundary_state} of $B_{\rho_1,\rho_2;2}$ vanish in all center-odd sectors, because $S^{(4)}_{2\mu}=0$ for odd $\mu$, whereas those of $\rho=1,3$ do not. The boundary entropies and the center selection rule are therefore consistent with the proposed QRW candidates, the vacuum label matching $B_{0,0;2}$, $B_{3,0;2}$ and the Fibonacci label $B_{1,0;2}$, $B_{2,0;2}$. Identifying the individual pull-backs requires further data, such as a mixed annulus with a known QRW boundary: the two nontrivial rotations of $\mathbb Z_3$ are conjugate under the quantum $\mathbb Z_2$ of the extension and so belong to a single orbit, whose canonical orbifold projection gives one image, whereas $B_{0,0;2}$ and $B_{3,0;2}$ have opposite nonzero coefficients in the $(1,1;0)$ sector. The construction also explains why the interface has no defect conformal manifold within the $\mathrm{SU}(2)$-preserving class~\cite{Antinucci2025,Chen2026a}: the Cartan zero modes of $(G_2)_1$ generate continuous families of interfaces that preserve only a $U(1)$ subgroup, and the $\mathrm{SU}(2)$-symmetric points on them are isolated, the analog of the RG angle of Ref.~\cite{Chen2026a}. For $k=7$ the phantom primary has conformal weight two and transforms under the diagonal $\mathrm{SU}(2)$ as $\tfrac72\otimes\tfrac12=3\oplus4$; both stress tensors are singlets, so this symmetry forbids direct mixing with either of them, and the extension must be analyzed separately~\cite{Furuta2025}. For $k\equiv1\pmod4$ it has half-integer spin and the extension is fermionic~\cite{FTZ2021}; we leave these cases for future work.

\section{Numerical details}
\label{app:numerics}
\paragraph*{Candidates.} Table~\ref{tab:results} lists the lowest levels of the undressed QRW representatives $B_{0,0;\rho}$ for $k=3$, i.e.\ $\rho_1=\rho_2=0$, obtained from Eq.~\eqref{eq:annulus}: in the coset vacuum sector the affine label obeys $\nu'\in\rho\otimes\rho$, while for general $\nu$ it occurs whenever some $\lambda\in\rho\otimes\rho$ has $N^{(k+1)\,\nu'}_{\lambda\nu}>0$; for $\rho=1$, for instance, $(0,0;2)\otimes4$ occurs through $\lambda=2$ although $4\notin1\otimes1$. These values are specific to the undressed representatives: away from the symmetry-selected orbit $\rho=\rho_\ast$ the coset-vacuum sector does not contain every even $\suk{k+1}$ module, and a non-invertible UV dressing adds sectors of positive coset weight that can nevertheless contain charged states of smaller total weight. For instance $B_{1,0;1}$ at $k=3$ contains $(2,0;2)\otimes4$ with multiplicity one, giving $h_2=1/15+1$ and $h_2/h_1=16/5$ instead of $4$, and $B_{1,0;2}$ at $k=5$ gives $h_3/h_1=43/7$ instead of $7$. The comparison in the main text is therefore with the representatives displayed, supplemented by the dressing test of Eq.~\eqref{eq:dressed_singlet}. For $\rho=1,3$ the lowest $S^z=2$ state of $B_{0,0;\rho}$ is the level-one descendant $J^+_{-1}$ of the spin-one primary, $h=4/3$, so that $h_2/h_1=4$; the lowest $S^z=2$ \emph{primary} lies higher, in $(0,0;2)\otimes4$ at $h=5/3$, and would give $5$. Since the lattice ground state of a given $S^z$ sector is the lowest state irrespective of whether it is primary, the relevant comparison is with $4$. For the factorizing representatives $\rho=0,4$ the labels are tied, the spectrum is that of two decoupled open segments, and the lowest levels are $0,1,2,3,4,7$ in units of $h_1=1$. For comparison, a homogeneous $\suk{1}$ ring gives $h_m/h_1=m^2$ (App.~\ref{app:homogeneous}).

\paragraph*{Method.} In Eq.~\eqref{eq:HWZW} each three-site term is labeled by its central site $i$, with site indices understood modulo $L$; $J_{3,i}=0$ for $i$ in the left half and $J_{3,i}=J_3^c$ for $i$ in the right half, so that the model is symmetric under the reflection of the ring that exchanges the two junctions, which therefore fold to the same boundary state. The energies are obtained by DMRG on periodic rings, targeting the lowest state of each magnetization sector separately, with bond dimensions up to $\chi=4000$ and a truncation error $\varepsilon_{\rm tru}=10^{-8}$; the second level of the $S^z=0$ sector at $L=40$ comes from a separate run targeting the two lowest states of that sector. All fits and stability measures are computed from the unrounded energies. The same parameters are used for the homogeneous rings of App.~\ref{app:homogeneous}, for the spin-$5/2$ chains of App.~\ref{app:su25}, and for the two robustness checks described below, in which the couplings or the segment lengths are changed. Periodic boundary conditions are costly for matrix-product states, so the accessible sizes are smaller than for open chains; two independent runs at $L=40$ give ground-state energies that differ by $5\times10^{-4}$ in units of the first gap, which sets the scale of the residual numerical uncertainty. The dressing test of Eq.~\eqref{eq:dressing_test} uses the two lowest $S^z=0$ levels of such a run: on the $L=40$ ring their difference is $0.99997$ times the $S^z=1$ gap. 

\paragraph*{Finite-size corrections.} We use $c+a/L^{b}$ as a phenomenological finite-size ansatz. A defect operator of dimension $\Delta$ added to the interface has a coupling of dimension $1-\Delta$ and can produce a relative correction $\propto L^{1-\Delta}$ to the level spacing, i.e.
\begin{equation}
  b=\Delta-1 ,
  \label{eq:bexp}
\end{equation}
provided that its first-order matrix elements in the states considered are nonzero and that its contribution does not cancel in the normalized gap; an irrelevant bulk operator of dimension $\Delta_{\rm bulk}$ can similarly produce $b=\Delta_{\rm bulk}-2$, and marginal bulk operators produce logarithms~\cite{Cardy1986log,AGSZ1989,chepiga2020floating}. The defect operators are the states of $Z_{BB}$, graded by $\mathrm{SU}(2)$ and by the center according to Eq.~\eqref{eq:centre_parity}, so the candidates can be listed. The only relevant singlet, $\phi_{1,3}\otimes\chi_0$ at $\Delta=2/3$ [the $q^{2/3}$ term of Eq.~\eqref{eq:ZA}], is center-odd; it also has vanishing diagonal matrix elements between the states realizing the magnetization minima in the equal-propagation-time annulus of $B_{0,0;2}$, which all lie in the coset-identity sector, because $(0,0;0)\times(0,0;0)=(0,0;0)$, so that at first order it cannot shift them, although at second order it would contribute $\propto L^{2/3}$. The lightest irrelevant singlet, the descendant $J_{-1}\!\cdot\!\Phi$ of the spin-one primary at $\Delta=4/3$, belongs to the coset-identity sector and is not removed by this selection rule. Nevertheless, because it is center-odd, its diagonal first-order matrix elements vanish in the center eigenstates of the unperturbed boundary theory, whatever the microscopic coupling, since $J\ket n=\eta_n\ket n$ and $JO_-J^{-1}=-O_-$ give $\bra n O_-\ket n=0$; the lowest state of each sector is nondegenerate apart from its $\mathrm{SU}(2)$ partners, so no degenerate perturbation theory is needed. In the absence of relevant degeneracies the leading effect of either center-odd operator from a single coupling is of second order, of relative order $L^{2(1-\Delta)}$: a growth $\propto L^{2/3}$ for $\Delta=2/3$ and a decay $\propto L^{-2/3}$ for $\Delta=4/3$, with possible mixed terms if several couplings are present; for the relevant coupling $u$ this perturbative scaling applies only while $|u|L^{1/3}\ll1$, before the crossover generated by $u$. As discussed in the main text, the center selection rule is established for the continuum interface, whereas the microscopic junction breaks one-site translation explicitly through the position-dependent $J_{3,i}$; for the lattice junction the data are compatible with proximity to the center-symmetric boundary over the accessible size range, but they do not separately determine the center-odd couplings or exclude a longer crossover scale. Among the center-even singlets, those at $\Delta=2$ are the two stress tensors $T_{\rm cos}$ and $T_{\suk{4}}$---$T_{\rm cos}$ does not shift the gaps between states that share the coset vacuum, and $T_{\suk{4}}$ rescales them by a common factor, so that neither changes the normalized gap---and the level-one singlet descendants of the two center-even copies of the spin-one primary, which lie in a nontrivial coset sector and have vanishing diagonal matrix elements; the next singlets are at $\Delta=3$. Bulk operators, such as the scalar descendant $J^a_{-1}\bar J^b_{-1}\Phi^{ab}_{j=1}$ of dimension $2+4/5$ in the $\suk{3}$ half, can add corrections of their own. The fitted exponents reported below should therefore be read as effective exponents of the ansatz rather than as identifications of specific operators.

\paragraph*{Extrapolation.} We fit $c+a/L^{b}$ with all three parameters free (scanning $b$ and solving linearly for $a$ and $c$), and estimate the uncertainty from leave-one-out resampling over the seven sizes $L=40,60,\dots,160$. The sectors behave differently. The $S^z=2$ data decrease monotonically, whereas the $S^z=5$ data are predominantly decreasing; in both sectors the fit is stable, giving $b\simeq1.1$ and $b\simeq2.1$ and
\begin{equation}
  R_2(L\!\to\!\infty)=3.139(9),\qquad R_5(L\!\to\!\infty)=18.66(9),
\end{equation}
where the quoted uncertainties are heuristic stability measures, the larger of the leave-one-out spread and half the range of the finite-size data; they do not include the systematic uncertainty associated with the extrapolation ansatz. For $S^z=3$ and $S^z=4$ the data are not monotonic---they pass through a shallow minimum near $L=80$ and rise again---and vary by only $0.005$ and $0.015$ over the whole range, i.e.\ by $0.07\%$ and $0.12\%$; the largest ring, $L=160$, continues the slow rise of the preceding sizes. The available sizes do not resolve a stable extrapolation there, and we report finite-size averages and half-ranges,
\begin{equation}
  \overline{R}_3=6.925(2),\qquad \overline{R}_4=12.199(8),
\end{equation}
as descriptive measures rather than as controlled estimates of the asymptotic values and their uncertainties; flatness over a finite window does not by itself prove that the asymptotic value has been reached. These lattice ratios are the quantities compared with $h_m/h_1$ in Table~\ref{tab:results}. For comparison, fixing $b\in\{1,\frac32,2\}$ and averaging gives $3.141(3)$, $6.926(1)$, $12.200(4)$, $18.64(2)$, and a logarithmic form $c+a/\ln(L/L_0)$ gives $3.10$, $6.93$, $12.21$, $18.2$; all procedures agree within $3\%$ and none of them changes the conclusions. Adding the ring with $L=160$ to the six smaller ones changes the finite-size averages for $S^z=3,4$ by $0.0004$ and $0.0013$, and the fitted limits for $S^z=2,5$, obtained from the unrounded energies with the same protocol, by $0.0008$ and $0.005$.

\paragraph*{Check at a second set of couplings.} The pattern $0\!:\!1\!:\!3\!:\!7\!:\!12\!:\!19$ is not tied to the couplings $J_2=0.1$, $J_3^c=0.05128$. Repeating the calculation with equal segment lengths at $J_2=0$, with $J_3=0.063$ on the $\suk{3}$ side, the $\suk{3}$ critical point of the model at this value of $J_2$~\cite{Michaud:2013nq,chepiga2020floating}, and $J_3=0$ on the other side, the pure spin-$3/2$ Heisenberg chain in the $\suk{1}$ phase, gives a similar pattern of normalized gaps over the sizes studied. The agreement supports the robustness of the observed
spectral pattern under this change of microscopic
couplings.

\begin{figure}[h]
\centering
\includegraphics[width=0.92\textwidth]{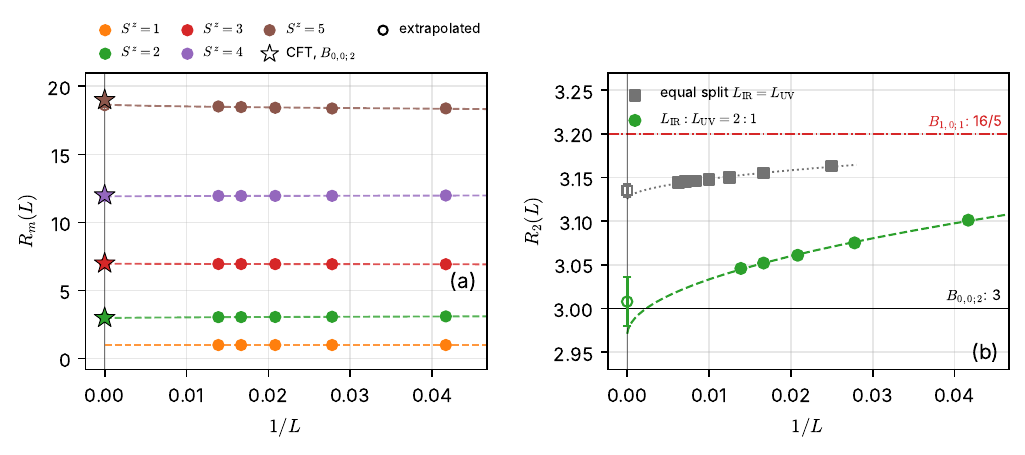}
\caption{Rings with $L_{\rm IR}:L_{\rm UV}=2:1$, chosen to reduce
the estimated propagation-time imbalance, at $L=24,36,48,60,72$. (a) Normalized gaps versus $1/L$, fits $c+a/L^{1/2}$ (dashed), the fixed-exponent extrapolations (open circles), and Eq.~\eqref{eq:hm_k3} (stars). (b) The discriminating sector $S^z=2$: the $2{:}1$ rings (circles) compared with the equal-split rings of Fig.~\ref{fig:defect_spectrum} (squares) and with the predictions $3$ of $B_{0,0;2}$ and $16/5$ of the dressed competitor $B_{1,0;1}$.}
\label{fig:matched}
\end{figure}

\paragraph*{Unequal velocities and a $2{:}1$ geometry.}
 Equation~\eqref{eq:ratio} is compared with Eq.~\eqref{eq:hm_k3} under the assumption of equal propagation times, $\tau=L_{\rm UV}/v_{\rm UV}=L_{\rm IR}/v_{\rm IR}$, in which case the spectrum is $E-E_0=\pi h/\tau$; for equal lengths this requires equal velocities. The bulk velocities of the two halves are different: $v_{\rm UV}=1.095$ at the $\suk{3}$ point~\cite{chepiga2020floating}, confirmed by our homogeneous UV rings (App.~\ref{app:homogeneous}), while the homogeneous $\suk{1}$ rings give an effective $v_{\rm IR}$ that rises from $2.05$ to $2.41$ over $L=40$--$160$, well described by $v^{\rm eff}_{\rm IR}(\ell)\simeq2.76-4.47/\sqrt\ell$. To reduce the propagation-time imbalance, we also study
rings with $L_{\rm IR}=2L_{\rm UV}$ and $L=24,36,48,60,72$.
Using the above phenomenological estimate gives
$\tau_{\rm IR}/\tau_{\rm UV}\simeq1.33,1.19,1.11,1.07,1.04$,
respectively, suggesting a closer approach to equal
propagation times. These estimates extrapolate the
homogeneous-ring fit below its fitted size range and do
not establish exact matching. With sites indexed modulo
$L$, the reflection $i\mapsto L_{\rm IR}-1-i$ still
exchanges the two junctions while preserving each segment,
so both junctions fold to the same boundary state.
The $2{:}1$ rings give
\begin{equation}
  R_m(L)=\begin{cases}
  3.101,\ 6.932,\ 11.981,\ 18.377 & (L=24),\\
  3.075,\ 6.936,\ 11.958,\ 18.374 & (L=36),\\
  3.061,\ 6.943,\ 11.954,\ 18.425 & (L=48),\\
  3.052,\ 6.949,\ 11.954,\ 18.478 & (L=60),\\
  3.046,\ 6.952,\ 11.953,\ 18.523 & (L=72),
  \end{cases}\qquad(m=2,\dots,5),
  \label{eq:matched}
\end{equation}
shown in Fig.~\ref{fig:matched}. Even at these smaller total sizes, the $2{:}1$ geometry
improves agreement in the discriminating sector $S^z=2$,
while all four displayed ratios remain within about
$3.4\%$ of Eq.~\eqref{eq:hm_k3}. For $S^z=2$, the equal-split extrapolation gives
approximately $3.14$, whereas the five $2{:}1$ data
points decrease monotonically and lie closer to $3$; the sectors $S^z=3,5$ rise slowly towards $7$ and $19$, and $S^z=4$ is flat. The fixed-exponent ensemble $b\in\{\tfrac12,1,\tfrac32\}$ gives
\begin{equation}
  R_m(L\!\to\!\infty)=3.008(28),\ \ 6.964(11),\ \ 11.930(14),\ \ 18.58(8)\qquad(m=2,\dots,5),
\end{equation}
where parentheses give the spread over the three exponents, and a free fit of $b$ to the $S^z=2$ data gives $b=0.78$ and $R_2=3.005$; leaving out any one size changes the $b=\tfrac12$ intercept by less than $0.01$. The value $3.008(28)$ singles out the symmetry-selected $3$ against the $16/5=3.2$ of the dressed competitor $B_{1,0;1}$, which the equal-split value $3.14$ could not clearly separate. The absolute gap is consistent with this picture: for equal propagation times $L\,\Delta E_1=\pi v_{\rm UV}\,(L/L_{\rm UV})\,h_1=\pi v_{\rm UV}\simeq3.44$, and the $2{:}1$ rings give $3.16$, $3.26$, $3.32$, $3.37$, $3.40$, approaching this value, whereas on equal halves the predictions from $\tau_{\rm UV}$ and $\tau_{\rm IR}$ differ by a factor of two. The remaining deviations, below $1\%$ in $S^z=3,4$ and $2\%$ in $S^z=5$, are compatible with the finite crossover length of the IR segment (App.~\ref{app:homogeneous}).

\section{Homogeneous-chain checks of the IR phase}
\label{app:homogeneous}
The RG-interface interpretation requires that the left half of the chain, $J_3=0$ at $J_2=0.1$, lies in the $\suk{1}$ phase, reached from the nearby $\suk{3}$ point. According to the phase diagram of the model~\cite{Michaud:2013nq,chepiga2020floating} and Haldane's conjecture for half-integer spins~\cite{Haldane1983,Affleck1987}, this point lies in the critical $\suk{1}$ phase at a distance $J_3^c\simeq0.051$ from the $\suk{3}$ line. The IR phase fixes the long-distance theory; whether a controlled $\suk{3}$ scaling window exists at short distances is a separate question. We test the IR identification directly on homogeneous rings with $J_1=1$, $J_2=0.1$, $J_3=0$ and $L=40$--$160$.

\paragraph*{Reference values.} On a homogeneous ring of the diagonal $\suk{k}$ model the lowest state with $S^z=J^3_0+\bar J^3_0=m$ minimizes $h+\bar h$ over the sectors $(j,j)$. For $m\le k$ it is the state with $J^3_0=\bar J^3_0=m/2$ in $j=m/2$, with $h+\bar h=m(m+2)/[2(k+2)]$, so that $h_m/h_1=m(m+2)/3$; for larger $m$ the lowest state is a current descendant. One finds
\begin{align}
  \suk{1}:&\quad \frac{h_m}{h_1}=m^2=0,\,1,\,4,\,9,\,16,\,25,\nonumber\\
  \suk{3}:&\quad \frac{h_m}{h_1}=0,\,1,\,\tfrac83,\,5,\,\tfrac{28}{3},\,\tfrac{43}{3},
  \label{eq:ring_ref}
\end{align}
where for $\suk{3}$ the $m=4$ and $m=5$ entries come from the descendants $J^+_{-1}\bar J^+_{-1}$ in the sector $(1,1)$ and $(J^+_{-1})^2(\bar J^+_{-1})^2$ in the sector $(\tfrac12,\tfrac12)$, respectively ($h+\bar h=14/5$ and $43/10$, with $h_1=3/10$). For $m\ge2$ the $\suk{3}$ values lie $33$--$44\%$ below the $\suk{1}$ ones, so in the scaling limit the normalized gaps identify the fixed point unambiguously.

\begin{figure}[h]
\centering
\includegraphics[width=0.6\textwidth]{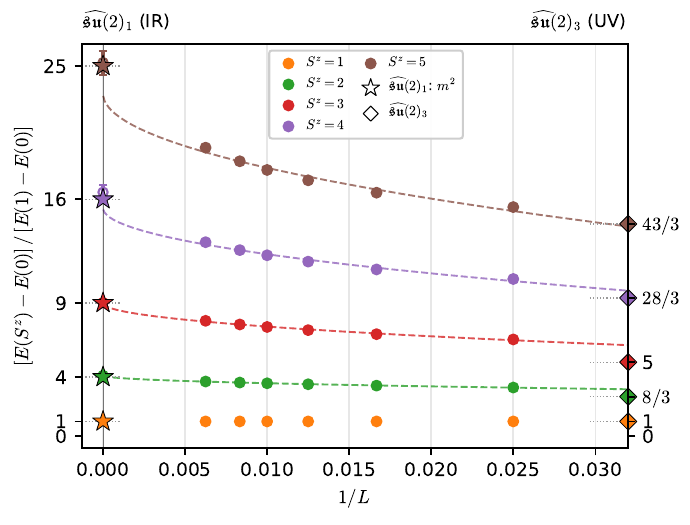}
\caption{Normalized gaps of the homogeneous spin-$3/2$ ring at $J_2=0.1$, $J_3=0$ versus $1/L$. The left axis is labeled by the $\suk{1}$ values $m^2$ (stars at $1/L=0$), the right axis by the $\suk{3}$ values of Eq.~\eqref{eq:ring_ref} (diamonds). Dashed lines are fits $c+a/L^{1/2}$ to all six sizes, continued slightly beyond the data as a visual guide only; that they run towards the $\suk{3}$ values at larger $1/L$ is suggestive but not evidence of a controlled crossover. Open circles are the extrapolated values, obtained from all sizes for $S^z=2,3$ and from the three largest for $S^z=4,5$, which is why for the latter two they do not coincide with the intercept of the curve shown. With increasing $L$ the data leave the $\suk{3}$ values and approach the $\suk{1}$ ones.}
\label{fig:homogeneous}
\end{figure}

\paragraph*{Homogeneous UV rings.} The same check at the UV point confirms that the right half of the interface ring sits at the $\suk{3}$ critical point. Homogeneous rings with $J_2=0.1$ and $J_{3}=J_3^c$ on every site give, for $L=80$ and $120$, the normalized gaps $(h_m/h_1)=(1,\,2.649,\,4.948,\,9.282,\,14.254)$ and $(1,\,2.684,\,5.052,\,9.440,\,14.501)$ for $S^z=1,\dots,5$, to be compared with the $\suk{3}$ values $1,\tfrac83,5,\tfrac{28}3,\tfrac{43}3$, with which they agree to $1$--$2\%$ with no extrapolation. The lowest triplet of these rings, $h+\bar h=3/10$, gives $v_{\rm UV}=1.14$ and $1.11$ at $L=80$ and $120$, approaching the value $1.095$ of Ref.~\cite{chepiga2020floating}, so that Eq.~\eqref{eq:HWZW} has the same normalization as that reference.

\paragraph*{Results.} Figure~\ref{fig:homogeneous} shows the data. The normalized gaps increase monotonically with $L$. At $L=40$ they are $3.29$, $6.53$, $10.6$, $15.5$ for $m=2,\dots,5$, already above the $\suk{3}$ values; at $L=160$ they are $3.69$, $7.79$, $13.1$, $19.5$, between $8\%$ ($m=2$) and $22\%$ ($m=5$) below the $\suk{1}$ values. The approach is slow, as expected for a crossover that ends at a fixed point with a marginally irrelevant current--current coupling~\cite{Cardy1986log,AGSZ1989}: the effective exponent, obtained from three consecutive sizes without assuming the limit, grows from near zero to about $0.5$ over the available range for $m=2$, while for $m=3,4,5$ it reaches only about $0.3$, $0.15$ and $0.03$ on the three largest sizes, so that no single power law describes the whole data set. We therefore use $b=1/2$, close to the largest effective exponent observed and a common phenomenological choice for such extrapolations, and extrapolate the sectors $S^z=4,5$---where the crossover is slowest and the smallest rings are furthest from the fixed point---from the three largest sizes only, the sectors $S^z=2,3$ from all six. The quoted value is the mean, and the quoted uncertainty the standard deviation, over an ensemble of unweighted least-squares intercepts: the fit to the full window at $b=1/2$, the leave-one-out fits to that window at $b=1/2$, and the fits to the full window at $b=0.4$ and $0.6$. This gives
\begin{equation}
  R_m(L\!\to\!\infty)=4.10(6),\ \ 9.01(20),\ \ 16.5(5),\ \ 25.2(8)\qquad(m=2,\dots,5),
\end{equation}
in agreement with $m^2=4,9,16,25$ within about $3\%$ in every sector. For reference, the single unweighted fit at $b=1/2$ to the same windows gives $4.092$, $8.986$, $16.434$, $25.167$; holding the windows fixed and varying the exponent gives $4.240$, $9.444$, $17.371$, $26.766$ at $b=0.4$ and $3.993$, $8.681$, $15.809$, $24.102$ at $b=0.6$. These spreads measure the sensitivity of the chosen fit family, not a statistical confidence interval. Using all six sizes for $S^z=4,5$ instead gives $15.4$ and $23.1$, below the reference values by $4$--$8\%$; the power-law fits with $b=1$ and the logarithmic fits lie on opposite sides of $m^2$, illustrating the extrapolation uncertainty of this phenomenological procedure.

The homogeneous-ring data thus support the identification of the left segment with the $\suk{1}$ IR phase. Their finite-size drift, away from the $\suk{3}$ values and towards the $\suk{1}$ ones, is consistent with, but does not by itself establish, a crossover from the nearby $\suk{3}$ critical point; establishing the crossover scale would require following it along several couplings approaching $J_3^c$. The drift occurs at sizes comparable to those used for the defect spectrum, which is a likely source of the residual percent-level deviations in Table~\ref{tab:results}.

\section{Spin-$5/2$ chain: $\suk{5}\to\suk{1}$}
\label{app:su25}
The spin-$5/2$ version of Eq.~\eqref{eq:HWZW} realizes $\suk{5}$ critical points~\cite{Chepiga:2022ciu}. We take $J_1=1$, $J_2=0.09$, $J_{3,i}=J_3^c=0.0232499$ on the right half, the $\suk{5}$ critical point of that model, and $J_{3,i}=0$ on the left half, and compute the lowest energy in each sector $S^z=0,\dots,5$ for $L=20,24,\dots,40$, all with even half-length. The undressed state $B_{0,0;3}$ predicts Eq.~\eqref{eq:hm_k5}, $h_m/h_1=0,1,3,6,11,17$, and $h_1=1/4$. The undressed representatives $B_{0,0;\rho}$ with the competing values of $\rho$ give $0,1,3,7,11,17$ ($\rho=2,4$), $0,1,5,9,13,17$ ($\rho=1,5$), and $0,1,2,3,4,5$ (factorizing $\rho=0,6$), so that the $S^z=2,3$ sectors distinguish these representatives; a homogeneous $\suk{5}$ ring would give $0,1,\frac83,5,8,\frac{35}{3}$.

\begin{figure}[h]
\centering
\includegraphics[width=0.92\textwidth]{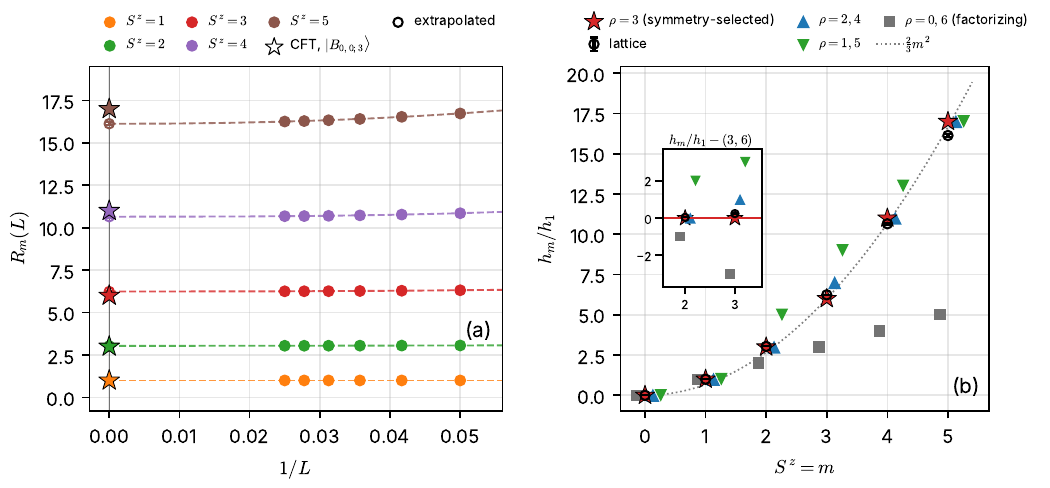}
\caption{Defect spectrum of the $\suk{5}\to\suk{1}$ interface on spin-$5/2$ rings with $L=20$--$40$, analyzed with the extrapolation procedure described in the text. (a) Normalized gaps versus $1/L$, fits $c+a/L^{b}$ with $b$ free, extrapolated values with their error bars (see text), and the prediction of $B_{0,0;3}$ (stars). (b) Extrapolated values compared with the undressed QRW representatives $B_{0,0;\rho}$: the symmetry-selected boundary label $\rho=3$, $\rho=2,4$, $\rho=1,5$, and the factorizing $\rho=0,6$, offset horizontally for visibility; dotted: $U(1)_{12}$ charging energy $\tfrac23m^2$. Inset: deviation from $(3,6)$ in the sectors $S^z=2,3$ that test the symmetry-selected module content.}
\label{fig:su25}
\end{figure}

\paragraph*{Results.} Figure~\ref{fig:su25} shows the data for $L=20,24,28,32,36,40$. The normalized gaps decrease monotonically and smoothly with $L$. Fitting $c+a/L^{b}$ with $b$ free, as in App.~\ref{app:numerics}, gives $b\simeq1.6,\,2.5,\,2.8,\,2.2$ for $S^z=2,\dots,5$ and
\begin{equation}
  R_m(L\!\to\!\infty)=3.036(2),\ \ 6.239(14),\ \ 10.65(5),\ \ 16.13(9)\qquad(m=2,\dots,5),
\end{equation}
where the fits use the unrounded energies and the quoted uncertainty is the larger of the leave-one-out spread and half the difference from the fixed-exponent ensemble $b\in\{1,\frac32,2\}$, which gives $3.033$, $6.21$, $10.56$, $15.94$; as in App.~\ref{app:numerics}, it does not include the systematic uncertainty of the ansatz. The sectors $S^z=2,3$ agree with the prediction $(3,6)$ to about $1.2\%$ and $4.0\%$ and favor $B_{0,0;3}$ among the undressed representatives displayed, within the equal-propagation-time comparison. The value at $S^z=2$ differs from the predictions of $B_{0,0;1}$, $B_{0,0;5}$, which give $5$, and of $B_{0,0;0}$, $B_{0,0;6}$, which give $2$; at $S^z=3$, the discriminating sector of Table~\ref{tab:general}, it lies within $4\%$ of $6$ and about $11\%$ from the $7$ of $B_{0,0;2}$, $B_{0,0;4}$ [Fig.~\ref{fig:su25}(b)]. These magnetization minima alone do not exclude all dressed QRW states; Eq.~\eqref{eq:dressed_singlet} provides an additional test of the non-invertibly dressed transmitting candidates. The sectors $S^z=4,5$ lie about $3.2\%$ and $5.1\%$ below the predicted $11$ and $17$, deviations of the same magnitude as those of the spin-$3/2$ chain in Table~\ref{tab:results}. The fitted powers are effective exponents of the finite-size ansatz and do not identify individual irrelevant operators. Part of the residual deviation may come from the finite crossover length of the IR half, which for $k=5$ starts from a multicritical point with two relevant singlet perturbations~\cite{Chepiga:2022ciu}; this crossover scale has not been determined independently and depends on the relevant couplings near the multicritical point. An independent finite-size test of the dressing is given below. \paragraph*{Dressing.} The lowest level of each magnetization sector does not distinguish the dressings. For $k=5$ the non-invertibly dressed states, $0<\rho_1<k$, carry an additional $\mathrm{SU}(2)$ singlet at $h=2/[(k+2)(k+3)]$, i.e.\ at $h/h_1=1/(k+2)=1/7$, so that the first excited level of the $S^z=0$ sector provides the same test as Eq.~\eqref{eq:dressing_test}: the undressed state $B_{0,0;3}$ predicts that this level coincides with the $S^z=1$ gap, whereas the non-invertible dressings predict one seventh of it. A separate run on the $L=20$ ring targeting the two lowest states of the $S^z=0$ sector gives a first excited level at $0.99998$ times the $S^z=1$ gap, i.e.\ equal to it within the run-to-run variation of the ground-state energy ($8\times10^{-4}$ of the gap). At $L=20$ the normalized level is therefore approximately one, seven times the non-invertibly dressed prediction. Under the equal-propagation-time comparison, this finite-size result favors the minimal-$g$ pair $B_{0,0;3}$ and $B_{5,0;3}$, which share this spectrum, over the non-invertibly dressed candidates, by the same factor $k+2$ by which their singlet lies below the first triplet; following the singlet levels with $L$ is needed to establish their IR ordering.

\end{document}